\documentclass[aps, prd, letterpaper, 12pt, nofootinbib, superscriptaddress, longbibliography, notitlepage]{revtex4-1}
\usepackage[utf8]{inputenc}
\usepackage{amsmath,amssymb,amsfonts}
\usepackage{mathrsfs}
\usepackage{color}
\usepackage{graphicx} 
\usepackage[section]{placeins}
\usepackage[colorlinks=true,linkcolor=blue,citecolor=blue]{hyperref}

\allowdisplaybreaks

\usepackage{tikz}
\usepackage{tkz-euclide}
\usetikzlibrary{decorations.pathmorphing}	
\tikzset{
    v/.style={decorate, decoration={snake, segment length=3mm, amplitude=0.75mm}, draw},
    f/.style={draw,decoration={markings,mark=at position #1 with {\arrow[very thick]{latex}}},postaction={decorate},node contents=#1},
    f/.default=.6,
    fb/.style={draw,decoration={markings,mark=at position #1 with {\arrowreversed[very thick]{latex}}},postaction={decorate},node contents=#1},
    fb/.default=.4,
    fnar/.style={draw},
    g/.style={decorate, draw,  decoration={coil,amplitude=3pt, segment length=3.5pt}},
    s/.style={dashed,draw, postaction={decorate},
        decoration={markings,mark=at position .55 with {\arrow[very thick]{latex}}}},
    sb/.style={dashed,draw, postaction={decorate},
        decoration={markings,mark=at position .55 with {\arrowreversed[draw=black,very thick]{latex}}}},
    snar/.style={dashed,draw,line width =1.25pt},
}
\usetikzlibrary{shapes}	
\tikzset{every picture/.style={line width=1}}

\definecolor{c1}{rgb}{0.121569, 0.466667, 0.705882}
\definecolor{c2}{rgb}{1., 0.498039, 0.054902}
\definecolor{c3}{rgb}{0.172549, 0.627451, 0.172549}
\definecolor{c4}{rgb}{0.839216, 0.152941, 0.156863}
\definecolor{c5}{rgb}{0.580392, 0.403922, 0.741176}
\definecolor{c6}{rgb}{0.54902, 0.337255, 0.294118}
\definecolor{c7}{rgb}{0.890196, 0.466667, 0.760784}
\definecolor{c8}{rgb}{0.498039, 0.498039, 0.498039}
\definecolor{c9}{rgb}{0.737255, 0.741176, 0.133333}
\definecolor{c10}{rgb}{0.0901961, 0.745098, 0.811765}

\begin{document}

\title{Probing Lepton Flavor Violation at \\ Circular Electron-Positron Colliders}

\author{Wolfgang~Altmannshofer}
\email{waltmann@ucsc.edu}
\affiliation{Department of Physics, University of California Santa Cruz, and
Santa Cruz Institute for Particle Physics, 1156 High St., Santa Cruz, CA 95064, USA}

\author{Pankaj~Munbodh}
\email{pmunbodh@ucsc.edu}
\affiliation{Department of Physics, University of California Santa Cruz, and
Santa Cruz Institute for Particle Physics, 1156 High St., Santa Cruz, CA 95064, USA}

\author{Talise~Oh}
\email{twoh@ucsc.edu}
\affiliation{Department of Physics, University of California Santa Cruz, and
Santa Cruz Institute for Particle Physics, 1156 High St., Santa Cruz, CA 95064, USA}

\begin{abstract}
Lepton flavor violation is one of the cleanest probes of physics beyond the standard model. In this work, we explore the sensitivity of the process $e^+ e^- \to \tau \mu$ to new physics above the TeV scale at the proposed circular electron-positron colliders FCC-ee and CEPC. We compute the $e^+ e^- \to \tau \mu$ cross-section in the Standard Model Effective Field Theory and assess the relevant backgrounds.
We compare our sensitivity projections to existing and expected constraints from tau decays and $Z$ decays and find that the future electron-positron colliders provide competitive probes of new physics. We highlight the complementarity of searches for resonant $e^+ e^- \to Z \to \tau \mu$ production on the $Z$ pole and searches for non-resonant $e^+ e^- \to \tau \mu$ at higher center-of-mass energies. 
\end{abstract}

\maketitle

\section{Introduction} \label{sec:intro}

In the Standard Model (SM), lepton flavor violation (LFV) is entirely absent at the tree-level, as the flavor eigenstates of the charged leptons coincide by definition with the corresponding mass eigenstates. LFV arises, in principle, at the loop-level, but it is highly suppressed by the neutrino masses and thus far below any foreseeable experimental sensitivities. Therefore, any observation of LFV at future experiments would be a smoking gun for new physics. For a recent review on LFV, see e.g.~\cite{Calibbi:2017uvl}.

No new physics state has been discovered after the Higgs in particle physics experiments. It is thus well motivated to consider the possibility that the new physics scale $\Lambda$ is substantially larger than the electroweak scale. A well-established model-independent framework to parametrize the effects of such heavy new physics is the Standard Model Effective Field Theory (SMEFT)~\cite{Grzadkowski:2010es}. The SMEFT contains 59 dimension-six operators (not counting flavor indices) of which a subset can contribute to LFV.
Many processes can be used to probe the LFV operators, including semi-leptonic and leptonic decays of vector mesons, taus, and muons at low energies, and decays of heavy particles such as the Higgs, $Z$ boson, or the top quark at higher energies~\cite{Altmannshofer:2022fvz}. Since one LFV operator can contribute to several of these processes simultaneously, information about the new physics can be accessed from a multitude of angles. 
Many studies point out the complementarity of the different probes and highlight the importance of treating LFV as a multifaceted feature of the new physics, see e.g.~\cite{Delepine:2001di, Gutsche:2011bi, Davidson:2012wn, Crivellin:2013hpa, Celis:2014asa, Calibbi:2021pyh, Calibbi:2022ddo}.

In this paper, we apply the SMEFT to study LFV in the process $e^+ e^- \to \tau \mu$ at future circular electron-positron colliders, namely the Future Circular Lepton Collider (FCC-ee)~\cite{FCC:2018evy, Bernardi:2022hny} and the Circular Electron Positron Collider (CEPC)~\cite{CEPCStudyGroup:2018ghi, CEPCPhysicsStudyGroup:2022uwl}. 
In the context of electron-positron collisions, various searches for LFV have been performed at LEP. DELPHI, ALEPH, L3, and OPAL all provide constraints on the LFV branching ratios of the $Z$ boson, $Z\to \ell_i \ell_j$ \cite{DELPHI:1992pgs, ALEPH:1991qhf, L3:1993dbo, OPAL:1995grn}. OPAL has also obtained constraints on the cross-sections of the non-resonant production of lepton flavor violating final states $e^+ e^- \to \ell_i \ell_j$ at energies above the $Z$ pole~\cite{OPAL:2001qhh}. Similar analyses can be envisioned at the FCC-ee and CEPC. Since the future colliders would run at much higher luminosities and span a greater energy range, we can expect a sizeable improvement in the sensitivity to the LFV operators. Indeed, reference \cite{Dam:2018rfz} estimated that the sensitivity to LFV $Z$ decays at the FCC-ee are up to four orders of magnitude better than at LEP, accounting for the huge expected luminosity and the improved detector resolution. Searches for LFV $Z$ decays involving taus at future $e^+ e^-$ colliders would probe new physics at a level that is comparable to that obtained from LFV tau decays~\cite{Calibbi:2021pyh}.

Since different SMEFT operators induced by the new physics have a characteristic scaling with the center-of-mass energy $\sqrt{s}$ at which they are probed, and FCC-ee and CEPC will run at a range of energies, we study $e^+ e^- \to \tau \mu$ both on the $Z$-pole and at higher energies. In particular, we explore the sensitivity of the $e^+ e^- \to \tau \mu$ searches to the new physics encapsulated by the SMEFT coefficients and point out the complementarity of the $Z$ pole and high energy runs. We also compare the sensitivity of $e^+ e^- \to \tau \mu$ to the sensitivity from low energy processes, rare tau decays in particular. Related studies of the process $e^+ e^ - \to \tau e$ at future lepton colliders (with a focus on the ILC) in the context of four-fermion contact interactions have been carried out in~\cite{Murakami:2014tna, Cho:2018mro, Etesami:2021hex}.

This paper is structured as follows: in section~\ref{sec:operators}, we introduce the SMEFT operators that contribute to the cross-section of $e^+ e^- \to \tau \mu$.
In section~\ref{sec:dilepton}, we give the full expression for the differential and total cross-section. Moreover, we discuss the effects of RGE evolution on the Wilson coefficients and the dependence of the coefficients on the center-of-mass energy.
In section~\ref{sec:constraints}, we review the existing constraints on the SMEFT operators coming from limits on low-energy tau decays at BaBar and Belle, on $Z$ decays at the LHC, and on $e^+ e^- \to \tau \mu $ production at LEP. 
In section~\ref{sec:sensitivities}, we assess the background and signal kinematic distributions at the future colliders, and compute the expected sensitivity of the FCC-ee and CEPC to the $e^+ e^- \to \tau \mu$ cross-section at different $\sqrt{s}$ energies.
In section~\ref{sec:results}, we discuss the constraints that can be expected from the future colliders and how they compare to those from low-energy tau decays, LEP, and the LHC.
We conclude in section~\ref{sec:conclusions}. Details about our treatment of the lepton flavor violating tau decays are given in appendix~\ref{app:tau_decays}.

\section{The Effective Field Theory Setup} \label{sec:operators}

There are three classes of SMEFT operators~\cite{Grzadkowski:2010es} that contribute at tree level to flavor-violating di-lepton production at electron-positron colliders
\begin{itemize}
\item[(i)] dipole operators
\end{itemize}
\begin{equation}
   (C_{eW})_{ij} (\bar \ell_i \sigma^{\mu\nu} e_j) \tau^I H W_{\mu\nu}^I ~,~~~  (C_{eB})_{ij} (\bar \ell_i \sigma^{\mu\nu} e_j) H B_{\mu\nu} ~, 
\end{equation}
\begin{itemize}
\item[(ii)] Higgs current operators
\end{itemize}
\begin{equation}
(C_{\varphi \ell}^{(1)})_{ij} (H^\dagger i \overset{\leftrightarrow}{D_\mu} H)(\bar \ell_i \gamma^\mu \ell_j) ~,~~~ (C_{\varphi \ell}^{(3)})_{ij} (H^\dagger i \overset{\leftrightarrow}{D^I_\mu} H)(\bar \ell_i \gamma^\mu \tau^I \ell_j) ~,~~~
(C_{\varphi e})_{ij} (H^\dagger i \overset{\leftrightarrow}{D_\mu} H)(\bar e_i \gamma^\mu e_j) ~,
\end{equation}
\begin{itemize}
\item[(iii)] four fermion contact interactions 
\end{itemize}
\begin{equation}
(C_{\ell \ell})_{ijkl} (\bar \ell_i \gamma^\mu \ell_j) (\bar \ell_k \gamma^\mu \ell_l) ~, ~~~ (C_{ee})_{ijkl} (\bar e_i \gamma^\mu e_j) (\bar e_k \gamma^\mu e_l) ~,~~~
(C_{\ell e})_{ijkl} (\bar \ell_i \gamma^\mu \ell_j) (\bar e_k \gamma^\mu e_l) ~.
\end{equation}
In these expressions, $\ell_i$ and $e_i$ are the three generations of lepton doublets and charged lepton singlets, $H$ is the Higgs doublet, and $W_{\mu\nu}^I$, $B_{\mu\nu}$ are the field strength tensors of the $SU(2)_L$ and $U(1)_Y$ gauge bosons. The Higgs currents are given by
\begin{equation}
H^\dagger i \overset{\leftrightarrow}{D_\mu} H = i \left( H^\dagger (D_\mu H) - (D_\mu H^\dagger) H \right) ~,~~
H^\dagger i \overset{\leftrightarrow}{D^I_\mu} H = i \left( H^\dagger \tau^I (D_\mu H) - (D_\mu H^\dagger) \tau^I H \right) ~,
\end{equation}
and $\tau^I$ are the Pauli matrices.

While SMEFT has become a standard framework for parameterizing heavy beyond SM physics, we find it convenient to work with a set of operators that is tailored to flavor-violating di-lepton production at electron-positron colliders. In the following, we focus on the operators relevant for $\tau^+ \mu^-$ production. The operators for $\mu^+ \tau^- $ production are simply the conjugate versions.

We prefer to formulate dipole operators in terms of the photon and $Z$ boson field strengths $F_{\mu \nu}$ and $Z_{\mu \nu}$
\begin{align}
    (C_\gamma^{LR})_{\mu\tau} & \frac{1}{\sqrt{2}} \frac{v}{\Lambda^2} (\bar \mu \sigma^{\alpha\beta} P_R\tau) F_{\alpha\beta} ~, \quad & (C_Z^{LR})_{\mu\tau} & \frac{1}{\sqrt{2}} \frac{v}{\Lambda^2} (\bar \mu \sigma^{\alpha\beta} P_R\tau) Z_{\alpha\beta} ~, \\
    (C_\gamma^{RL})_{\mu\tau} & \frac{1}{\sqrt{2}} \frac{v}{\Lambda^2} (\bar \mu \sigma^{\alpha\beta} P_L \tau) F_{\alpha\beta} ~, \quad & (C_Z^{RL})_{\mu\tau} & \frac{1}{\sqrt{2}} \frac{v}{\Lambda^2} (\bar \mu \sigma^{\alpha\beta} P_L \tau) Z_{\alpha\beta} ~,
\end{align}
we consider the flavor violating $Z$ couplings provided by the Higgs current operators
\begin{align}
    (C_Z^{LL})_{\mu\tau} & \frac{v^2}{2\Lambda^2} (\bar \mu \gamma^{\alpha} P_L \tau) \frac{g}{c_W} Z_{\alpha} ~, \quad & (C_Z^{RR})_{\mu\tau} & \frac{v^2}{2\Lambda^2} (\bar \mu \gamma^{\alpha} P_R \tau) \frac{g}{c_W} Z_{\alpha} ~,
\end{align}
and also consider the four fermion operators
\begin{align}
    (C_V^{LL})_{\mu\tau} & \frac{1}{\Lambda^2} (\bar e \gamma_\alpha P_L e)(\bar \mu \gamma^{\alpha} P_L \tau) ~, \quad & (C_V^{RR})_{\mu\tau} & \frac{1}{\Lambda^2} (\bar e \gamma_\alpha P_R e)(\bar \mu \gamma^{\alpha} P_R \tau) ~, \\
    (C_V^{LR})_{\mu\tau} & \frac{1}{\Lambda^2} (\bar e \gamma_\alpha P_L e)(\bar \mu \gamma^{\alpha} P_R \tau) ~, \quad & (C_V^{RL})_{\mu\tau} & \frac{1}{\Lambda^2} (\bar e \gamma_\alpha P_R e)(\bar \mu \gamma^{\alpha} P_L \tau) ~, \\
    (C_S^{LR})_{\mu\tau} & \frac{1}{\Lambda^2} (\bar e P_L e)(\bar \mu P_R \tau) ~, \quad & (C_S^{RL})_{\mu\tau} & \frac{1}{\Lambda^2} (\bar e P_R e)(\bar \mu P_L \tau) ~.
\end{align}
The Wilson coefficients of these operators are given by linear combinations of SMEFT coefficients. We find
\begin{align}
\label{eq:CgammaLR}
(C_{\gamma}^{LR})_{\mu \tau} &= c_W (C_{eB})_{\mu\tau} - s_W (C_{eW})_{\mu \tau} ~, \quad & (C_Z^{LR})_{\mu \tau} &= - c_W (C_{eW})_{\mu \tau} - s_W (C_{eB})_{\mu \tau} ~, \\
(C_{\gamma}^{RL})_{\mu\tau} &= c_W (C_{eB})^*_{\tau\mu} - s_W (C_{eW})^*_{\tau\mu} ~, \quad & (C_Z^{RL})_{\mu \tau} &= - c_W (C_{eW})^*_{\tau\mu} - s_W (C_{eB})^*_{\tau\mu} ~, \\
(C_Z^{LL})_{\mu \tau} &= (C_{\varphi \ell}^{(1)})_{\mu \tau} + (C_{\varphi \ell}^{(3)})_{\mu \tau}~, \quad &  (C_Z^{RR})_{\mu \tau} &= (C_{\varphi e})_{\mu \tau} ~, \\
(C_V^{LR})_{\mu \tau} &= (C_{\ell e})_{ee\mu\tau} ~, \quad & (C_V^{RL})_{\mu \tau} &= (C_{\ell e})_{\mu\tau ee} ~,  \\
(C_S^{LR})_{\mu \tau} &= - 2(C_{\ell e})_{\mu ee \tau} ~, \quad & (C_S^{RL})_{\mu \tau} &= - 2 (C_{\ell e})_{e \tau \mu e} ~,
\end{align}
\begin{align}
(C_V^{LL})_{\mu \tau} &= (C_{\ell\ell})_{ee\mu\tau} + (C_{\ell\ell})_{\mu\tau ee} + (C_{\ell\ell})_{e\tau \mu e} + (C_{\ell\ell})_{\mu e e\tau} ~, \\
\label{eq:CVRR}
(C_V^{RR})_{\mu \tau} &= (C_{ee})_{ee\mu\tau} + (C_{ee})_{\mu\tau ee} + (C_{ee})_{e\tau\mu e} + (C_{ee})_{\mu ee \tau} ~.
\end{align}
The expressions for the $e^+ e^- \to \tau \mu$ cross-section take a particularly compact form when using these Wilson coefficients. However, when presenting the result of our numerical analysis, we decide to show constraints on SMEFT coefficients in order to facilitate comparison with the literature.

Generalizing the above setup to $e^+ e^- \to \mu e$  and $e^+ e^- \to \tau e$ requires care, as in those cases some of the above operators are related by Fierz transformations. We leave a detailed discussion of $e^+ e^- \to \mu e$  and $e^+ e^- \to \tau e$ for future work.

Based on dimensional analysis, we can immediately determine the behavior of the $e^+ e^- \to \tau \mu$ cross-section at a large (squared) center of mass energy, $s$. If the cross-section is induced by four fermion operators, it will increase linearly with $s$. Dipole operators give a contribution that is approximately constant at large $s$, while the Higgs current operators give a contribution that falls like $1/s$. 
On top of that, the contributions of operators that contain the $Z$ boson will be resonantly enhanced on the $Z$ pole.

\section{Flavor Violating Di-Lepton Production at Electron Positron Colliders} \label{sec:dilepton}
\begin{figure}[tb]
\centering
\includegraphics[width=0.42\linewidth]{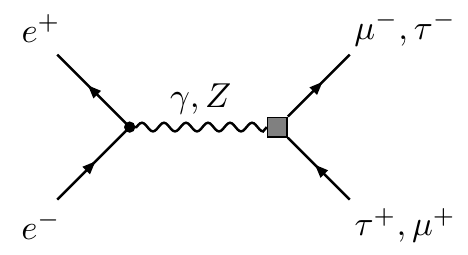} \qquad 
\includegraphics[width=0.3\linewidth]{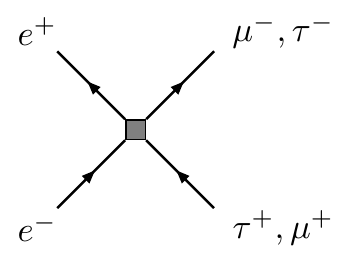}
\caption{Feynman diagrams for $e^+ e^- \to \tau \mu$ from lepton flavor violating $Z$ boson and photon couplings (left) and four fermion contact interactions (right).}
\label{fig:diagrams}
\end{figure}

We concentrate on the process $e^+e^- \to \tau \mu$ for which flavor violating $Z$ boson and photon interactions contribute only in the $s$-channel, see Fig.~\ref{fig:diagrams} for the corresponding Feynman diagrams. For the related processes $e^+e^- \to \tau e$ and $e^+e^- \to \mu e$ there are also $t$-channel contributions. For those processes, we expect qualitatively similar results but leave a detailed discussion for future work. The main advantage of only $s$-channel contributions is that the angular distribution has a particularly simple form. In fact, the differential cross sections of the processes $e^+e^- \to \tau^+ \mu^-$ and $e^+e^- \to \mu^+ \tau^-$ can be written as
\begin{equation}
    \frac{d \sigma(e^+e^- \to \tau^+ \mu^-)}{d\cos\theta} = \frac{d \sigma}{d\cos\theta} = \frac{1}{32\pi} \frac{m_Z^2}{\Lambda^4} \Big[ I_0(s) (1 + \cos^2\theta ) + I_1(s) \cos\theta + I_2(s) \sin^2\theta \Big] ~,
\end{equation}
\begin{equation}
    \frac{d \sigma(e^+e^- \to \mu^+ \tau^-)}{d\cos\theta} = \frac{d \bar\sigma}{d\cos\theta} = \frac{1}{32\pi} \frac{m_Z^2}{\Lambda^4} \Big[ \bar I_0(s) (1 + \cos^2\theta ) + \bar I_1(s) \cos\theta + \bar I_2(s) \sin^2\theta \Big] ~.
\end{equation}
We define $\theta$ to be the angle between the incoming electron and outgoing muon or the incoming positron and the outgoing anti-muon, respectively. The coefficients $I_i$ and $\bar I_i$ are functions of the (squared) center-of-mass energy $s$ and are observables. It is convenient to rearrange the six observables and bring the differential cross sections into the following form
\begin{eqnarray}
    \frac{1}{\sigma_\text{tot}} \frac{d (\sigma + \bar \sigma)}{d\cos\theta} &=& \frac{3}{8} (1 - F_D) (1 + \cos^2\theta ) + A_\text{FB} \cos\theta + \frac{3}{4} F_D \sin^2\theta  ~, \\
    \frac{1}{\sigma_\text{tot}} \frac{d (\sigma - \bar \sigma)}{d\cos\theta} &=& \frac{3}{8} (A^\text{CP} - F_D^\text{CP}) (1 + \cos^2\theta ) + A_\text{FB}^\text{CP} \cos\theta + \frac{3}{4} F_D^\text{CP} \sin^2\theta ~,
\end{eqnarray}
where $\sigma_\text{tot}= \sigma(e^+e^- \to \tau \mu)$ is the total cross section
\begin{equation}
\label{eq:SMEFT_cross_sec}
\sigma_\text{tot} = \sigma(e^+e^- \to \tau^+ \mu^-) + \sigma(e^+e^- \to \mu^+ \tau^-) = \frac{1}{24\pi} \frac{m_Z^2}{\Lambda^4} \Big( 2(I_0 + \bar I_0) + I_2 + \bar I_2 \Big) ~,
\end{equation}
and $A^\text{CP}$ is the direct CP asymmetry
\begin{equation}
A^\text{CP} = \frac{\sigma(e^+e^- \to \tau^+ \mu^-) - \sigma(e^+e^- \to \mu^+ \tau^-)}{\sigma(e^+e^- \to \tau^+ \mu^-) + \sigma(e^+e^- \to \mu^+ \tau^-)} = \frac{2(I_0 - \bar I_0) + I_2 - \bar I_2}{2(I_0 + \bar I_0) + I_2 + \bar I_2} ~.
\end{equation}
The angular observables $A_\text{FB}$ and $F_D$ are the CP averaged forward-backward asymmetry and the CP averaged dipole fraction
\begin{equation}
A_\text{FB} = \frac{3(I_1 + \bar I_1)}{4 ( 2(I_0 + \bar I_0) + I_2 + \bar I_2)} ~, \quad F_D = \frac{I_2 + \bar I_2}{2(I_0 + \bar I_0) + I_2 + \bar I_2} ~,
\end{equation}
and $A_\text{FB}^\text{CP}$ and $F_D^\text{CP}$ are the corresponding CP asymmetries
\begin{equation}
A_\text{FB}^\text{CP} = \frac{3(I_1 - \bar I_1)}{4 ( 2(I_0 + \bar I_0) + I_2 + \bar I_2)} ~, \quad F_D^\text{CP} = \frac{I_2 - \bar I_2}{2(I_0 + \bar I_0) + I_2 + \bar I_2} ~.
\end{equation}

For the $I_i$ and $\bar I_i$ coefficients, we find the expressions below. Note that we drop the flavor indices on the Wilson coefficients to improve the readability of the expressions.
\begin{multline}
I_0 + \bar I_0 = \frac{s}{2 m_Z^2} \left( |C_V^{LL}|^2 + |C_V^{RR}|^2 + |C_V^{LR}|^2 + |C_V^{RL}|^2 + \frac{1}{2}|C_S^{LR}|^2 + \frac{1}{2}|C_S^{RL}|^2 \right) \\
+ \frac{s^2}{(s - m_Z^2)^2 + \Gamma_Z^2 m_Z^2} \Bigg[ \frac{m_Z^2}{2 s} \Big( |C_Z^{LL}|^2 + |C_Z^{RR}|^2 \Big)(1 - 4 s_W^2 + 8 s_W^4) + \left( 1 - \frac{m_Z^2}{s} \right) \\
\times  \left( \text{Re}\Big( C_V^{LL} C_Z^{LL*} + C_V^{LR} C_Z^{RR*}\Big) (1 - 2 s_W^2) - \text{Re}\Big( C_V^{RL} C_Z^{LL*} + C_V^{RR} C_Z^{RR*}\Big) 2 s_W^2 \right) \Bigg] ~,
\end{multline}
\begin{multline}
I_0 - \bar I_0 = - \frac{s \Gamma_Z m_Z }{(s - m_Z^2)^2 + \Gamma_Z^2 m_Z^2} \\
\times  \Big( \text{Im}\left( C_V^{LL} C_Z^{LL*} + C_V^{LR} C_Z^{RR*}\Big) (1 - 2 s_W^2) - \text{Im}\Big( C_V^{RL} C_Z^{LL*} + C_V^{RR} C_Z^{RR*}\Big) 2 s_W^2 \right) ~,
\end{multline}
\begin{multline}
I_1 + \bar I_1 = \frac{s}{m_Z^2} \left( |C_V^{LL}|^2 + |C_V^{RR}|^2 - |C_V^{LR}|^2 - |C_V^{RL}|^2\right) \\
+ \frac{s^2}{(s - m_Z^2)^2 + \Gamma_Z^2 m_Z^2} \Bigg[ \frac{m_Z^2}{s} \Big( |C_Z^{LL}|^2 - |C_Z^{RR}|^2 \Big)(1 - 4 s_W^2) + 2 \left( 1 - \frac{m_Z^2}{s} \right) \\
\times  \left( \text{Re}\Big( C_V^{LL} C_Z^{LL*} - C_V^{LR} C_Z^{RR*}\Big) (1 - 2 s_W^2) + \text{Re}\Big(C_V^{RL} C_Z^{LL*} - C_V^{RR} C_Z^{RR*}\Big) 2 s_W^2 \right) \Bigg] ~,
\end{multline}
\begin{multline}
I_1 - \bar I_1 = - \frac{2 s \Gamma_Z m_Z }{(s - m_Z^2)^2 + \Gamma_Z^2 m_Z^2} \\
\times  \Big( \text{Im}\left( C_V^{LL} C_Z^{LL*} - C_V^{LR} C_Z^{RR*}\Big) (1 - 2 s_W^2) + \text{Im}\Big( C_V^{RL} C_Z^{LL*} - C_V^{RR} C_Z^{RR*}\Big) 2 s_W^2 \right) ~,
\end{multline}
\begin{multline}
I_2 + \bar I_2 = 8 \left( |C_\gamma^{LR}|^2 + |C_\gamma^{RL}|^2\right) c_W^2 s_W^2 + \frac{s}{4 m_Z^2} \left( |C_S^{LR}|^2 + |C_S^{RL}|^2\right) \\
+ \frac{s^2}{(s - m_Z^2)^2 + \Gamma_Z^2 m_Z^2} \Bigg[ \Big( |C_Z^{LR}|^2 + |C_Z^{RL}|^2 \Big)(1 - 4 s_W^2 + 8 s_W^4) \\ 
+ \left( 1 - \frac{m_Z^2}{s} \right) \text{Re}\Big( C_\gamma^{LR} C_Z^{LR*} + C_\gamma^{RL} C_Z^{RL*}\Big) 4 c_W s_W (1 - 4 s_W^2)  \Bigg] ~,
\end{multline}
\begin{equation}
I_2 - \bar I_2 = - \frac{s \Gamma_Z m_Z}{(s - m_Z^2)^2 + \Gamma_Z^2 m_Z^2}  \text{Im}\Big( C_\gamma^{LR} C_Z^{LR*} + C_\gamma^{RL} C_Z^{RL*}\Big) 4 c_W s_W (1 - 4 s_W^2) ~.
\end{equation}
We note that the CP asymmetries are all proportional to the width of the $Z$ boson and, therefore, typically negligible. The above expressions clearly reflect the dependence on the center-of-mass energy (squared) $s$ anticipated in section~\ref{sec:operators}.

The first step at future $e^+ e^-$ colliders is to search for a non-zero rate of $e^+ e^- \to \tau \mu$. Barring (presumably tiny) efficiency and acceptance corrections, this can be done independently of the angular distribution of the new physics signal and possible CP-violating effects. 
If a $e^+ e^- \to \tau \mu$ signal were to be established at a future collider, the angular observables and the CP asymmetries provide diagnostics about the underlying new physics and may, at least in principle, allow us to disentangle contributions from operators with different chirality structure. 

Before moving on, let us comment on the possible effects of renormalization group (RG) evolution. The Wilson coefficients that induce  $e^+ e^- \to \tau \mu$ production should be evaluated at a scale that approximately corresponds to the center-of-mass energy of the collider. As we will see in section~\ref{sec:results}, the typical new physics scales that can be probed at future circular electron-positron colliders are around 10 TeV, while the center-of-mass energy of the colliders is in the few hundred GeV range. The RG evolution between the new physics scale $\Lambda$ and the center of mass energy $\sqrt{s}$ typically gives corrections of the order of $\sim \alpha\log(\Lambda^2/s)$, which is at most a percent level effect and thus negligible. In our numerical analysis, we include the full set of 1-loop RG effects from electroweak gauge couplings and the top Yukawa coupling in a leading log approximation, implementing the results from~\cite{Jenkins:2013wua, Alonso:2013hga}\footnote{Note that some results from these papers are updated at~\cite{Manohars_webpage}.}. (We note that tools that can perform the running numerically are also available~\cite{Aebischer:2018bkb, Fuentes-Martin:2020zaz, DiNoi:2022ejg}.) There are two cases in which RG running might, in principle, become relevant.

First, starting with four fermion contact interactions at the high new physics scale, RG evolution can generate Wilson coefficients of the operators that contain $Z$ bosons. The loop suppression of these Wilson coefficients can be partly compensated by the resonant enhancement of the cross-section by $\sim m_Z^2/\Gamma_Z^2$ for colliders running on the $Z$-pole. Second, starting with the Higgs current operators at the high new physics scale, RG evolution can generate four fermion contact interactions. In that case, the loop suppression can be partly compensated by the relative energy enhancement of the cross-section by $\sim s^2/m_Z^4$ at colliders running at large center-of-mass energy. 
In practice, we find that those effects hardly play any role.

\begin{figure}[tbh]
\centering
\includegraphics[width=0.8\textwidth]{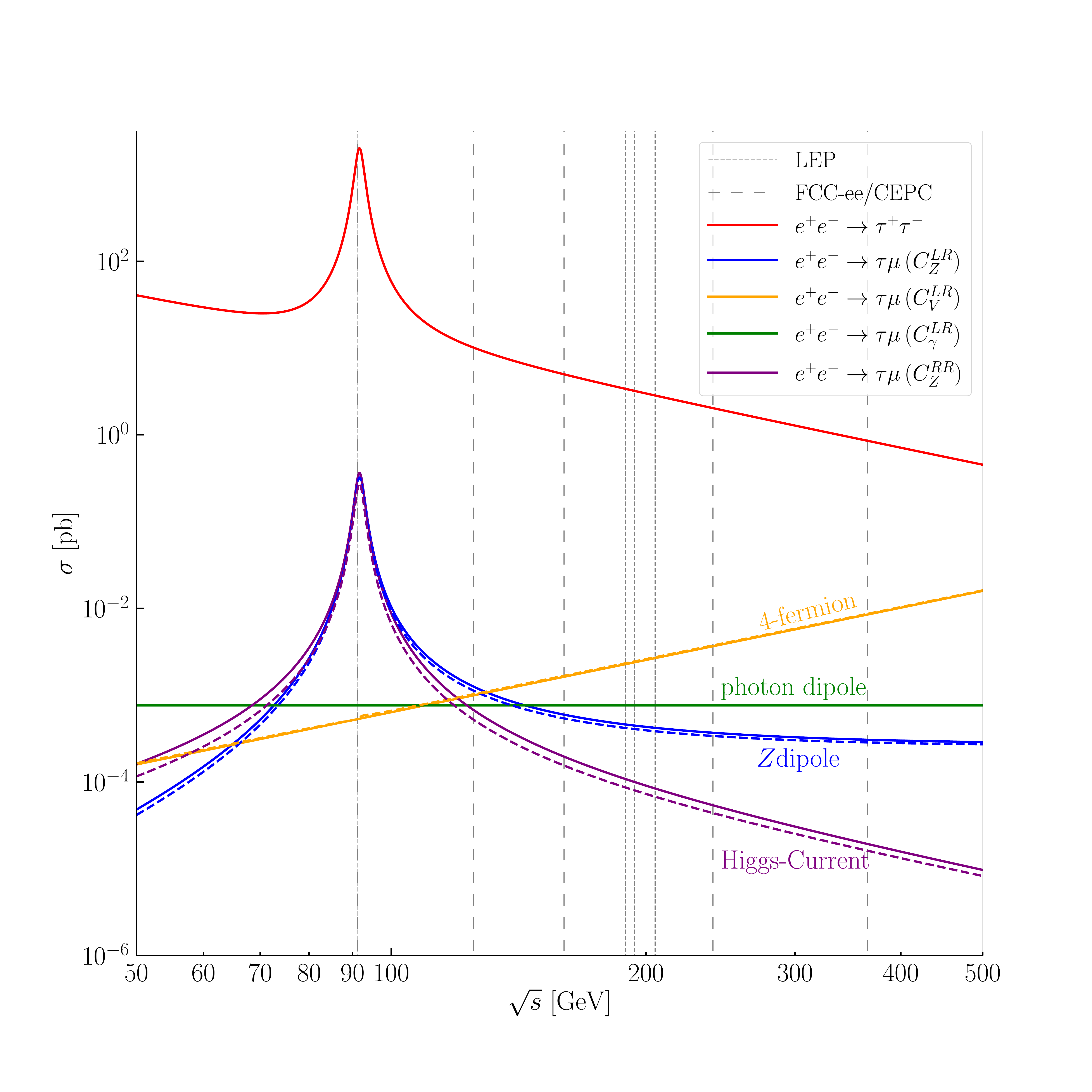}
\caption{The cross-section of the lepton flavor violating process $e^+ e^- \to \tau \mu$ as function of the center of mass energy $\sqrt{s}$. The blue, orange, green, and purple lines show the cross-section generated by the Wilson coefficients $C_Z^{LR}$, $C_V^{LR}$, $C_\gamma^{LR}$, and $C_Z^{RR}$, respectively. The new physics scale is set to $\Lambda = 3~\text{ TeV}$, and the Wilson coefficients are set to 1, one at a time. The solid lines show the tree-level result. The dashed lines include 1-loop renormalization group running. The red line shows the SM cross-section of $e^+e^- \to \tau^+ \tau^-$ for comparison. The vertical gray dotted (dashed) lines show the center of mass energies of the LEP (FCC-ee/CEPC).}
\label{fig:cross_sec}
\end{figure}

In figure~\ref{fig:cross_sec}, we show the total $e^+ e^- \to \tau \mu$ cross section, as a function of the center of mass energy $\sqrt{s}$. The blue, orange, green, and purple lines are obtained by setting the Wilson coefficients $C_Z^{LR}$, $C_V^{LR}$, $C_\gamma^{LR}$, and $C_Z^{RR}$ to unity, one at a time. For definiteness, the new physics scale is set to $\Lambda = 3~\text{TeV}$. The dashed lines include 1-loop renormalization group running, while the solid lines are the tree-level result. The effect of the RG running hardly makes a difference. For comparison, the red line shows the SM cross-section of $e^+e^- \to \tau^+ \tau^-$.
We note that in the calculation of the cross-sections, we do not include the effect of initial state radiation of photons. Close to the $Z$-pole, initial state radiation can have a noticeable impact on the cross-section~\cite{Greco:1980mh, ALEPH:2005ab}. A detailed evaluation of this effect is left for future work.

The vertical dotted and dashed lines in figure~\ref{fig:cross_sec} show the center-of-mass energies of LEP and proposed circular $e^+ e^-$ colliders, respectively. Combining the information from searches at several $\sqrt s$ values promises to provide complementary information about the new physics, as different operators provide the dominant new physics cross-section at different center-of-mass energies.

\section{Existing Constraints} \label{sec:constraints}

The operators that can lead to $e^+ e^- \to \tau \mu$ at future colliders are already probed by various existing experimental results, including lepton flavor violating decays of taus, lepton flavor violating decays of the $Z$ boson, as well as LEP searches for $\tau\mu$ production. 
We discuss these constraints in the following sections~\ref{sec:tau_decays},~\ref{sec:Z_decay}, and~\ref{sec:LEP}. 

\subsection{Lepton flavor violating tau decays} \label{sec:tau_decays}

The most important constraints are obtained from radiative, leptonic, and semileptonic flavor violating tau decays like $\tau\to \mu \gamma$, $\tau \to 3 \mu$, $\tau \to \mu e^+ e^-$, $\tau \to \mu \pi$, $\tau \to \mu \rho$, and $\tau \to \mu \phi$. All of the operators introduced in section~\ref{sec:operators} are bounded by one or more of these tau decays.

Details on how we calculate the tau branching ratios are given in appendix~\ref{app:tau_decays}. The relevant Feynman diagrams can be found in figure~\ref{fig:diagrams_tau_decays}. 

On the experimental side, the most stringent bounds on the lepton flavor violating tau decays have been established at BaBar~\cite{BaBar:2006jhm} and Belle~\cite{Hayasaka:2010np, Belle:2011ogy, Belle:2021ysv} (see also~\cite{Belle:2007cio, BaBar:2009wtb,  BaBar:2009hkt, BaBar:2010axs, LHCb:2014kws, ATLAS:2016jts, CMS:2020kwy})
\begin{eqnarray}
\text{BR}(\tau^- \to \mu^- e^+ e^-) &<& 1.8 \times 10^{-8} ~~ @ ~90\%~ \text{C.L.} ~, \\
\text{BR}(\tau^- \to \mu^- \mu^+ \mu^-) &<& 2.1 \times 10^{-8} ~~ @ ~90\%~ \text{C.L.} ~, \\
\text{BR}(\tau^- \to \mu^- \gamma) &<& 4.2 \times 10^{-8} ~~ @ ~90\%~ \text{C.L.} ~, \\
\text{BR}(\tau^- \to \mu^- \pi^0) &<& 1.1 \times 10^{-7} ~~ @ ~90\%~ \text{C.L.} ~, \\
\text{BR}(\tau^- \to \mu^- \rho^0) &<& 1.2 \times 10^{-8} ~~ @ ~90\%~ \text{C.L.} ~, \\
\text{BR}(\tau^- \to \mu^- \phi) &<& 8.4 \times 10^{-8} ~~ @ ~90\%~ \text{C.L.} ~.
\end{eqnarray}
At Belle II, the sensitivities will improve significantly. Assuming that the searches can be kept background-free, future constraints are expected to be approximately a factor 50 better~\cite{Belle-II:2018jsg}
\begin{eqnarray}
\text{BR}(\tau^- \to \mu^- e^+ e^-) &<& 3 \times 10^{-10} ~~ @ ~90\%~ \text{C.L.} ~, \\
\text{BR}(\tau^- \to \mu^- \mu^+ \mu^-) &<& 4 \times 10^{-10} ~~ @ ~90\%~ \text{C.L.} ~, \\
\text{BR}(\tau^- \to \mu^- \gamma) &<& 1 \times 10^{-9} ~~ @ ~90\%~ \text{C.L.} ~, \\
\text{BR}(\tau^- \to \mu^- \pi^0) &<& 5 \times 10^{-10} ~~ @ ~90\%~ \text{C.L.} ~, \\
\text{BR}(\tau^- \to \mu^- \rho^0) &<& 2 \times 10^{-10} ~~ @ ~90\%~ \text{C.L.} ~, \\
\text{BR}(\tau^- \to \mu^- \phi) &<& 1.5 \times 10^{-9} ~~ @ ~90\%~ \text{C.L.} ~.
\end{eqnarray}
The future lepton colliders FCC-ee and CEPC will have sensitivities that are comparable to Belle II, or slightly better~\cite{Dam:2018rfz, Dam:2021ibi, CEPCPhysicsStudyGroup:2022uwl, Bernardi:2022hny}.

\subsection{Lepton flavor violating decays of the Z boson} \label{sec:Z_decay}

The operators that contain the $Z$ boson can be directly probed by searches for lepton flavor violating decays of the $Z$. We calculate the branching ratio $\text{BR}(Z\to\tau \mu)$ in our setup and find it convenient to normalize by the measured (flavor averaged) branching ratio of the $Z$ boson to charged leptons
\begin{equation}
\frac{\text{BR}(Z \to \tau \mu)}{\text{BR}(Z \to \ell^+ \ell^-)} = \frac{2 v^4}{\Lambda^4} \frac{1}{1 - 4 s_W^2 + 8 s_W^4} \Big( |C_Z^{LL}|^2 + |C_Z^{RR}|^2 + |C_Z^{LR}|^2 + |C_Z^{RL}|^2 \Big) ~,
\end{equation}
with BR$(Z \to \ell^+ \ell^-) \simeq 3.3658\%$~\cite{ParticleDataGroup:2022pth}.
In the above expression, the Wilson coefficients are evaluated at the electroweak scale $m_Z$. Our result agrees with the one given in the appendix of~\cite{Crivellin:2013hpa}. 

The strongest experimental constraints on the $Z$ decays come from the LHC. In particular, ATLAS has established the bound~\cite{ATLAS:2020zlz, ATLAS:2021bdj}
\begin{equation}
\text{BR}(Z \to \tau \mu) < 6.5 \times 10^{-6} ~~ @ ~95\%~ \text{C.L.} ~.
\end{equation}
Since the ATLAS searches have backgrounds, the sensitivity is expected to improve with the square root of the luminosity, implying a sensitivity of $\sim 10^{-6}$ after the high-luminosity LHC~\cite{Altmannshofer:2022fvz}.

\subsection{\boldmath $e^+ e^- \to \tau \mu$ at LEP} \label{sec:LEP}

Also LEP provides constraints on the process $e^+ e^- \to \tau\mu$ which are straightforward to interpret in our formalism. Searches for non-standard $\mu \tau$ production have been carried out both on the $Z$-pole and at high energy runs~\cite{ALEPH:1991qhf, DELPHI:1992pgs, L3:1993dbo, OPAL:1995grn, OPAL:2001qhh}. The OPAL analysis~\cite{OPAL:2001qhh} provides directly 95\% confidence level bounds on the $e^+ e^- \to \tau \mu$ cross-section at three center-of-mass energies
\begin{eqnarray}
\label{eq:LEP_OPAL1}
 \sigma(e^+e^- \to \mu \tau) < 115~\text{fb} ~~~&& @~ \sqrt{s} = 189~\text{GeV} ~, \\
 \sigma(e^+e^- \to \mu \tau) < 116~\text{fb} ~~~&& @~ \sqrt{s} \simeq 194~\text{GeV} ~, \\
 \label{eq:LEP_OPAL3}
 \sigma(e^+e^- \to \mu \tau) < 64~\text{fb} ~~~&& @~ \sqrt{s} \simeq 205~\text{GeV} ~,
\end{eqnarray}
where in the latter two cases, we averaged over the energy ranges quoted in~\cite{OPAL:2001qhh}.

At the $Z$-pole, the non-observation of $\tau \mu$ events is usually interpreted as a bound on the $Z \to \tau\mu $ branching ratio. The most stringent one is obtained by OPAL, BR$(Z \to \tau \mu) < 1.7 \times 10^{-5}$~\cite{OPAL:1995grn}. We re-interpret this result as a generic constraint on $\tau \mu$ production and derive a constraint on the $e^+ e^- \to \tau \mu$ cross section using the quoted limit on the number of non-standard $\tau \mu$ events in the fiducial signal region, $N_{\tau \mu} < 9.9$ at 95\% confidence level. In terms of the $e^+e^- \to \tau \mu$ cross-section, the number of events is given by
\begin{equation} \label{eq:Ntaumu}
    N_{\tau \mu} = \mathcal L_\text{int} \times \kappa_{\tau\mu} \times \epsilon_{\tau} \times \epsilon_{\mu} \times \int d\cos\theta  ~ \frac{d\sigma(e^+ e^- \to \tau \mu)}{d\cos\theta} ~,
\end{equation}
where $\kappa_{\tau\mu} \simeq 87.6\%$ is the preselection efficiency and $\epsilon_\mu = 55.1\%$, $\epsilon_\tau = 44.7\%$ are the lepton identification efficiencies. The angular integration is bounded by $|\cos\theta| < 0.68$~\cite{OPAL:1995grn}. 

The total integrated luminosity used in the OPAL analysis is $\mathcal L_\text{int} = 129$~pb$^{-1}$. However, not all the data was taken directly on the $Z$-pole. Based on the details on the LEP data taking periods given in~\cite{ALEPH:2005ab}, we interpret the total integrated luminosity of 129~pb$^{-1}$ as being composed of $\sim 105$~pb$^{-1}$ on the $Z$-pole, $\sqrt{s} = m_Z$, and $\sim 12$~pb$^{-1}$ each at $\sqrt{s} = 89.4$~GeV and $93.0$~GeV. This has not an entirely negligible effect because the cross-section close to the $Z$-pole depends sensitively on $\sqrt{s}$ for some of the new physics operators. The product of integrated luminosity and cross-section that enters equation~\eqref{eq:Ntaumu} should therefore be interpreted as 
\begin{equation}
 \mathcal L_\text{int} \int d\sigma \simeq 105~\text{pb}^{-1} \int d\sigma(m_Z) + 12~\text{pb}^{-1}\left( \int d\sigma(89.4~\text{GeV}) + \int d\sigma(93.0~\text{GeV}) \right) ~.
\end{equation}
As a cross-check, we used this prescription to reproduce the bound on BR$(Z \to \tau \mu)$ quoted in~\cite{OPAL:1995grn}.
Taking into account that SM production of $Z$ bosons on the $Z$-pole is subject to large QED corrections from photon radiation (reference~\cite{ALEPH:2005ab} states that the leading order cross-section is $\sim 36\%$ larger than the measured one), we reproduce the bound within $5\%$. 

\begin{figure}[tb]
\centering
\includegraphics[width=0.6\textwidth]{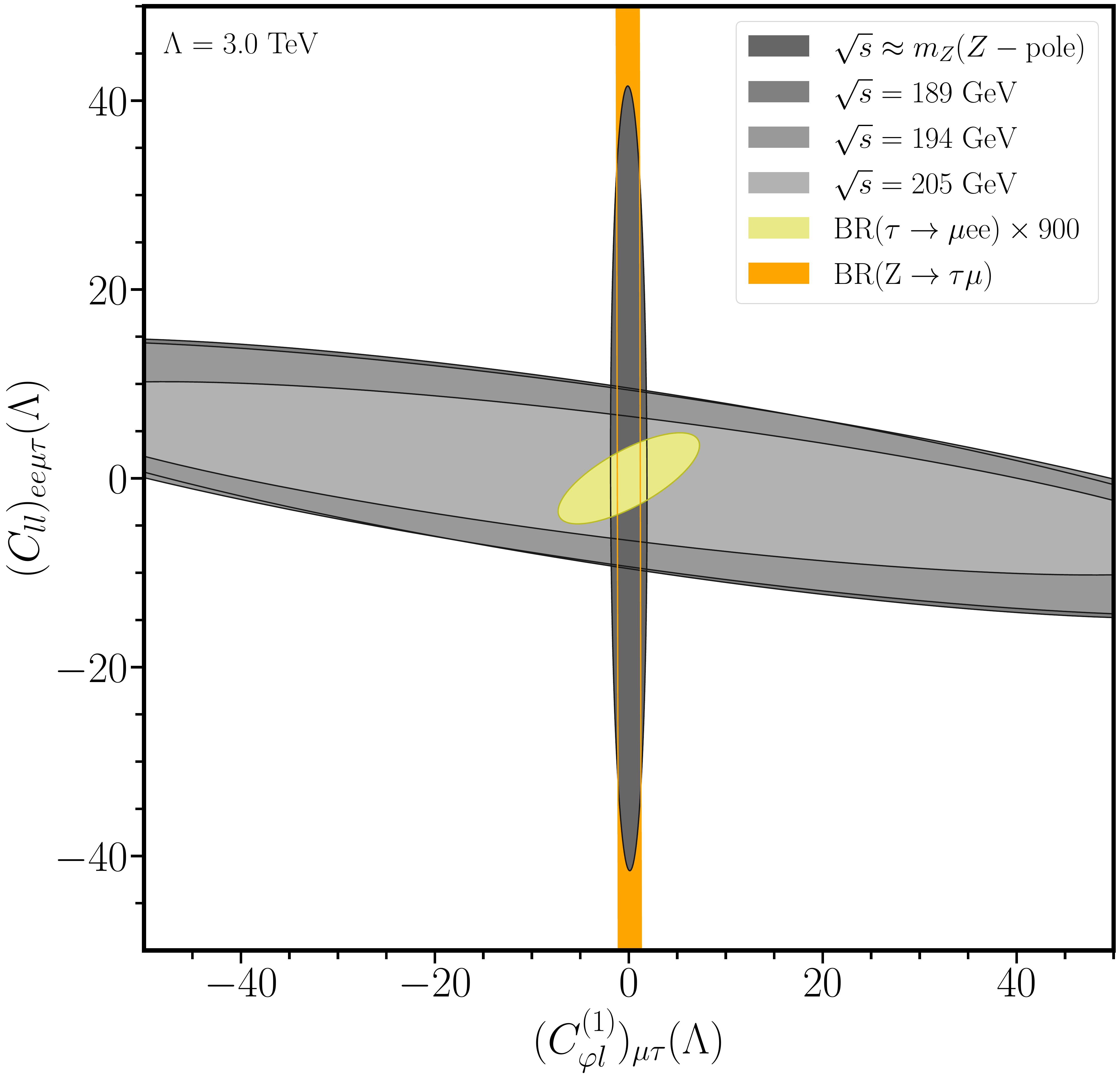}
\caption{$2\sigma$ constraints in the plane of the SMEFT Wilson coefficients $(C_{ll})_{ee\mu\tau}$ vs. $(C_{\varphi l}^{(1)})_{\mu\tau}$ from existing experimental results, in particular $e^+ e^- \to \tau \mu$ at LEP (gray), $Z \to \tau \mu$ from the LHC (orange), and $\tau \to \mu ee$ from the $B$ factories (yellow). The new physics scale is set to $\Lambda = 3$~TeV. The constraint from the tau decays is inflated by a factor of 900 for better visibility. This corresponds to a factor of 30 for the two Wilson coefficients.} 
\label{fig:Wilson_current}
\end{figure}

To illustrate the existing constraints on the lepton flavor violating operators, we show in figure~\ref{fig:Wilson_current} the bounds from $e^+ e^- \to \tau \mu$ at LEP (dark gray for the $Z$-pole results; lighter gray for the high-energy results), $Z \to \tau \mu$ from the LHC (orange), and the lepton flavor violating tau decays from the $B$ factories (yellow) in the two-dimensional plane of the SMEFT Wilson coefficients $(C_{ll})_{ee\mu\tau}$ vs. $(C_{\varphi l}^{(1)})_{\mu\tau}$. For definiteness, the new physics scale is set to $\Lambda = 3$~TeV. 

The $Z \to \tau \mu$ decay constrains the Higgs current coefficient $(C_{\varphi l}^{(1)})_{\mu\tau}$, but does not give any appreciable bound on the four fermion contact interaction $(C_{ll})_{ee\mu\tau}$ as it does not contribute to the $Z$ decay at tree level. Similarly, the $e^+ e^- \to \tau \mu$ process on the $Z$-pole mainly constrains the Higgs current interaction. On the other hand, the large $\sqrt{s}$ searches for $e^+ e^- \to \tau \mu$ at LEP give complementary information and constrain mainly the contact interaction. 
The $\tau \to \mu ee$ decay can constrain both considered Wilson coefficients simultaneously.
Note that the constraint from the tau decays is inflated in the figure by a factor of 900 for better visibility. This corresponds to a factor of 30 for the two Wilson coefficients. In fact, the current bound from the tau decays is much stronger than the combination of the $Z$ decays and $e^+ e^- \to \tau \mu$. 

In the following, we will see that the high expected sensitivity of future $e^+ e^-$ colliders can change this picture significantly.

\section{Expected Sensitivities at Future Circular Electron Positron Colliders} \label{sec:sensitivities}

To estimate the sensitivity to $e^+ e^- \to \tau \mu$ at future $e^+ e^-$ colliders, we follow a procedure analogous to the one proposed in~\cite{Dam:2018rfz} for the search for $Z \to \tau \mu$ at FCC-ee. We assess the relevant sources of background and obtain expected constraints on the $e^+ e^- \to \tau \mu$ cross-section for the various proposed runs at FCC-ee and CEPC.

\subsection{Signal and background kinematics} \label{sec:backgrounds}

The $e^+ e^- \to \tau \mu$ process is characterized by a tau in one hemisphere of the event, recoiling against a muon that has a momentum equal to the beam momentum in the other hemisphere. The analysis strategy proposed in~\cite{Dam:2018rfz} focuses on well reconstructable exclusive tau decays with more than one hadron in the final state like $\tau \to \rho \nu \to 2 \pi \nu $ and $\tau \to 3 \pi \nu$. This choice minimizes the probability of misidentifying a muon as a tau and thus reduces background from $e^+ e^- \to \mu^+ \mu^-$ to a negligible level. The remaining background sources are
\begin{itemize}
\item[(i)] $e^+ e^- \to \tau^+ \tau^- \to \tau_\text{had} \mu \nu \nu$ where one of the taus decays hadronically and the other decays to a muon, 
\item[(ii)] $e^+ e^- \to W^+ W^- \to \tau_\text{had} \nu \mu \nu$ where one of the $W$ bosons decays to a tau and the other to a muon,
\item[(iii)] $e^+ e^- \to W^+ W^- \to \tau^+ \nu \tau^- \nu$ or $e^+ e^- \to ZZ \to \tau^+ \tau^- \nu\nu$ with one of the taus decaying hadronically and the other to a muon. 
\end{itemize}
As we will see, all these backgrounds can be controlled by imposing cuts on the muon momentum, which we parameterize by the variable
\begin{equation} 
x = \frac{p_\mu}{p_\text{beam}} ~,
\end{equation}
where $p_\mu$ is the absolute value of the muon momentum and 
$p_\text{beam}$ is the absolute value of the beam momentum. For the signal, we expect that the muon momentum is equal to the beam momentum $x = 1$, while for the background part of the beam momentum is carried by undetected neutrinos and thus $x <1$. 

In the case of the background process (ii) $e^+ e^- \to W^+ W^- \to \tau_\text{had} \nu \mu \nu$ with two on-shell $W$ bosons, we find that the muon momentum is bounded by
\begin{equation}
x < \frac{1}{2} \left (1+ \sqrt{1-\frac{4 m_W^2}{s}} \right ) < 1 ~. 
\end{equation}
Given the precisely known beam energy and the expected excellent momentum resolution of the FCC-ee and CEPC detectors~\cite{CEPCStudyGroup:2018rmc, CEPCStudyGroup:2018ghi, FCC:2018evy, Blondel:2019jmp,  Gao:2022lew} this background can therefore be entirely removed by a cut on $x$. We have checked explicitly using MadGraph5 aMC@NLO~\cite{Alwall:2014hca}
that also the background from $e^+ e^- \to W^+ W^- \to \tau_\text{had} \nu \mu \nu$ with off-shell $W$ bosons is completely negligible.
We conclude that the same is true for the background processes (iii) $e^+ e^- \to W^+ W^- \to \tau^+ \nu \tau^- \nu$ and $e^+ e^- \to ZZ \to \tau^+ \tau^- \nu\nu$. 

For the background process (i) $e^+ e^- \to \tau^+ \tau^- \to \tau_\text{had} \mu \nu \nu$, the endpoint of the muon momentum distribution coincides with the beam momentum. In this case, a cut on $x \gtrsim 1$ retains an $\mathcal O(1)$ portion of the signal and removes the major part of the background. The remaining amount of background is determined by the beam energy spread and the muon momentum resolution of the detector.
Below we discuss in detail how to obtain the corresponding expected numbers of signal and background events.

\subsection{Signal and background event numbers} \label{sec:event_numbers}

The number of signal and background events can be written in the following way
\begin{align}
\label{eq:Nsig}
N_\text{sig} &= \sigma(e^+e^- \to \tau \mu) \times \text{BR}(\tau \to \text{had.}) \times \mathcal L_\text{int} \times \epsilon^{x_c}_\text{sig} \times \epsilon ~, \\
\label{eq:Nbkg}
 N_\text{bkg} &= \sigma(e^+e^- \to \tau^+ \tau^-) \times 2\times \text{BR}(\tau \to \text{had.}) \times \text{BR}(\tau \to \mu\bar\nu\nu) \times \mathcal L_\text{int} \times \epsilon^{x_c}_\text{bkg} \times \epsilon  ~.
 \end{align}
We denote with $\text{BR}(\tau \to \text{had.})$ the branching ratio of the tau into well-reconstructable hadronic final states. We will approximate this branching ratio as $\text{BR}(\tau \to \text{had.}) \simeq \text{BR}(\tau \to 4\pi\nu) + \text{BR}(\tau \to 3 \pi \nu) + \text{BR}(\tau \to 2\pi \nu) \simeq 47.53 \% $~\cite{ParticleDataGroup:2022pth}. We also use $\text{BR}(\tau \to \mu\bar\nu\nu) \simeq 17.39 \%$~\cite{ParticleDataGroup:2022pth}. As in~\cite{Dam:2018rfz}, we include a signal efficiency of $\epsilon \simeq 25\%$. The efficiencies $\epsilon^{x_c}_\text{bkg}$ and $\epsilon^{x_c}_\text{sig}$ are due to the cut on the muon momentum $x > x_c$, to be discussed below. The signal cross section $\sigma(e^+e^- \to \tau \mu)$ was given in Eq. \eqref{eq:SMEFT_cross_sec} in section~\ref{sec:dilepton}, while the $e^+ e^- \to \tau^+ \tau^-$ cross section in the SM can be written as
\begin{equation}
\label{eq:diff_cross_sec}
 \frac{d\sigma(e^+ e^- \to \tau^+ \tau^-)}{d\cos\theta} = \frac{3}{8} \sigma(e^+ e^- \to \tau^+ \tau^-) \left( 1 + \cos^2\theta + \frac{8}{3} A_\text{FB} \cos\theta \right) ~, 
\end{equation}
with the total cross-section
\begin{multline}
\label{eq:SM_cross_sec}
 \sigma(e^+ e^- \to \tau^+ \tau^-) = \frac{4 \alpha^2 \pi}{3 s} \Bigg[ 1 + \frac{s^2}{(s-m_Z^2)^2 + \Gamma_Z^2 m_Z^2} \Bigg( \frac{(1-4 s_W^2 + 8 s_W^4)^2}{64 s_W^4 c_W^4} \\
 + \frac{(1- 4 s_W^2)^2}{8 s_W^2 c_W^2} \left( 1 - \frac{m_Z^2}{s}\right) \Bigg) \Bigg] ~, 
\end{multline}
and the forward-backward asymmetry 
\begin{equation}
 A_\text{FB} = - \frac{3}{4} \frac{(1-4s_W^2)^2 + 8 s_W^2 c_W^2 (1-\frac{m_Z^2}{s})}{(1-4s_W^2+8s_W^4)^2+8s_W^2c_W^2(1-4s_W^2)^2(1-\frac{m_Z^2}{s})+64s_W^4c_W^4(1-\frac{m_Z^2}{s})^2} ~. 
\end{equation}
Due to the detector layout, a cut on the angle $\cos\theta$ will, in principle, need to be imposed (c.f. the discussion of the OPAL limit in section~\ref{sec:LEP}, where $|\cos\theta| < 0.68$). For simplicity, we do not include such a cut, both for signal and background. Such a cut could be easily implemented by using the differential cross-sections instead of the total cross-section and numerically integrating over the appropriate range of $\cos\theta$. We expect that a cut of that type has no significant impact on our final results.

The efficiencies $\epsilon^{x_c}_\text{bkg}$ and $\epsilon^{x_c}_\text{sig}$ are determined by the muon momentum distribution of the background and signal, which are smeared out due to the beam energy spread and the momentum resolution of the detector. We have 
\begin{equation}
    \epsilon^{x_c}_\text{bkg/sig} = \int_{x_c}^\infty dx \, \int dy\, \left(\frac{1}{\sigma} \frac{d\sigma}{dy} \right)_\text{bkg/sig} \frac{1}{\sqrt{2\pi} \delta x} \exp\left(-\frac{(x-y)^2}{2\delta x^2}\right) ~.
\end{equation}
To determine the smearing of $x$, we combine the uncertainty of the center-of-mass energy $\delta \sqrt{s}$ and the momentum resolution of the detector $\delta p_T$ in quadrature
\begin{equation}
\left(\frac{\delta x}{x}\right)^2 = \left(\frac{\delta \sqrt{s}}{\sqrt{s}}\right)^2 + \left(\frac{\delta p_T}{p_T}\right)^2 ~. 
\end{equation}
The muon momentum distributions of the signal and background are given by~\cite{Hagiwara:1989fn, Dam:2018rfz}
\begin{align}
\left(\frac{1}{\sigma}\frac{d\sigma}{dy}\right)_\text{sig} &= \delta(1-y) ~, \\
\left(\frac{1}{\sigma}\frac{d\sigma}{dy}\right)_\text{bkg} &= \left\{ \begin{array}{cl}
\frac{1}{3} \left [ (5 - 9 y^2 + 4 y^3) + P_\tau (1 - 9 y^2 + 8 y^3)\right ]  & ,~ \text{for} ~ 0 \leq y \leq 1 ~, \\
0 & ,~\text{otherwise} ~.
\end{array} \right.
\end{align}
The expression of the background distribution depends on $P_\tau$, the polarization asymmetry of the taus in $e^+ e^- \to \tau^+ \tau^-$. We find that it is given by
\begin{equation}
    P_\tau = \frac{-(1-4s_W^2)(1-4s_W^2 + 8 s_W^4 +8 s_W^2 c_W^2 (1-\frac{m_Z^2}{s}))}{(1-4s_W^2+8s_W^4)^2+8s_W^2c_W^2(1-4s_W^2)^2(1-\frac{m_Z^2}{s})+64s_W^4c_W^4(1-\frac{m_Z^2}{s})^2} ~. 
\end{equation} 
These are all the ingredients we need to determine the sensitivities at future $e^+ e^-$ colliders.

\subsection{Sensitivities at FCC-ee and CEPC} \label{sec:sensitivity_numerics}

To estimate the sensitivities at FCC-ee and CEPC to the $e^+ e^- \to \tau \mu$ cross-section, we impose the simple criterion
\begin{equation}
 N_\text{sig} \geq 2 \sqrt{N_\text{bkg} + N_\text{sig}} ~.
\end{equation}
In the limit of $N_\text{bkg} \gg N_\text{sig}$, this corresponds to a number of signal events larger than the $2\sigma$ statistical uncertainty on the background. In the absence of backgrounds $N_\text{bkg} \ll N_\text{sig}$, the criterion corresponds to $N_\text{sig} \geq 4$. 

{\setlength{\tabcolsep}{8pt}
\begin{table}[tb] 
\centering
\begin{tabular}{c c c c c c c}
\hline\hline
$\sqrt{s}$ [GeV]& $\mathcal L_{\text{int}}$ [ab$^{-1}$] & $\frac{\delta \sqrt{s}}{\sqrt{s}}$ [10$^{-3}$] & $\frac{\delta p_T}{p_T}$ [10$^{-3}$] & $\epsilon^{x_c}_\text{bkg}$ [10$^{-6}$] & $N_\text{bkg}$ & $\sigma$ [ab]\\ [0.5ex]
\hline
91.2 ($Z$-pole) & 75 & 0.93 & 1.35 & 1.55 & 9700 $\pm$ 100 & 45\\
87.7 (off-peak) & 37.5 & 0.93 & 1.33 & 1.46 & 520 $\pm$ 20 & 21 \\ 
93.9 (off-peak) & 37.5 & 0.93 & 1.37 & 1.59 & 930 $\pm$ 30 & 28 \\
125 ($H$) & 20 & 0.03 & 1.60 & 1.44 & 12 $\pm$ 3 & 8 \\
160 ($WW$)&  12 & 0.93 & 1.89 & 2.44 & 6 $\pm$ 2 & 10\\
240 ($ZH$) &  5 & 1.17 & 2.60 & 4.39 & 2 $\pm$ 1 & 18 \\
365 ($t\bar{t}$\,) &  1.5 & 1.32 & 3.78 & 8.61 & 0.5 $\pm$ 0.7 & 50 \\
\hline \hline 
\end{tabular} 
\caption{The center-of-mass energy, integrated luminosity, beam energy spread, and momentum resolution for FCC-ee~\cite{FCC:2018evy, Blondel:2019jmp, dEnterria:2021xij, Bernardi:2022hny}. The last three columns show the background efficiency, the expected number of background events with $1\sigma$ statistical uncertainty, and the derived sensitivity to the $e^+ e^- \to \tau \mu$ cross-section at 95\% confidence level.}
\label{table:info_FCC-ee}
\end{table}}
{\setlength{\tabcolsep}{8pt}
\begin{table}[tb] 
\centering
\begin{tabular}{c c c c c c c}
\hline\hline
$\sqrt{s}$ [GeV]& $\mathcal L_{\text{int}}$ [ab$^{-1}$] & $\frac{\delta \sqrt{s}}{\sqrt{s}}$ [10$^{-3}$] & $\frac{\delta p_T}{p_T}$ [10$^{-3}$] & $\epsilon^{x_c}_\text{bkg}$ [10$^{-6}$] & $N_\text{bkg}$ & $\sigma$ [ab]\\ [0.5ex]
\hline
91.2 ($Z$-pole) & 50 & 0.92 & 1.35 & 1.53 & 6400 $\pm$ 80 & 55\\
87.7 (off-peak) & 25 & 0.92 & 1.33 & 1.46 & 350 $\pm$ 20 & 27 \\ 
93.9 (off-peak) & 25 & 0.92 & 1.37 & 1.59 & 620 $\pm$ 25 & 35 \\
160 ($WW$)&  6 & 0.99 & 1.89 & 2.49 & 3 $\pm$ 2 & 17 \\
240 ($ZH$) &  20 & 1.20 & 2.60 & 4.42 & 7 $\pm$ 3 & 6.6 \\
360 ($t\bar{t}$\,) &  1 & 1.41 & 3.74 & 8.61 & 0.3 $\pm$ 0.5 & 72 \\
\hline\hline
\end{tabular} 
\caption{The center-of-mass energy, integrated luminosity, beam energy spread, and momentum resolution for CEPC~\cite{CEPCStudyGroup:2018rmc, CEPCStudyGroup:2018ghi, Gao:2022lew, CEPCPhysicsStudyGroup:2022uwl}. The last three columns show the background efficiency, the expected number of background events with $1\sigma$ statistical uncertainty, and the derived sensitivity to the $e^+ e^- \to \tau \mu$ cross-section at 95\% confidence level.}
\label{table:info_CEPC}
\end{table}}

The uncertainties on the center-of-mass energy $\delta \sqrt{s}$ can be determined from the total energy spreads in collision listed for FCC-ee in~\cite{Blondel:2019jmp} and for CEPC in~\cite{CEPCStudyGroup:2018rmc, Gao:2022lew} at various center-of-mass energies. 
The goal of a possible Higgs pole run, $\sqrt{s} = m_H$, is to determine the coupling of the Higgs to electrons by measuring the s-channel Higgs production.
Due to the very narrow width of the Higgs, this requires exquisite precision on the beam energy of the order of the Higgs width or better. On the Higgs pole, we use thus a benchmark value of $\delta\sqrt{s} = \Gamma_H \simeq 4.1$~MeV, as was also done in Ref.~\cite{dEnterria:2021xij}. Because of the finite momentum resolution of the detector, the precise value of $\delta\sqrt{s}$ on the Higgs pole has little impact on our final results.

For the momentum resolution of the detector we use $\delta p_T/p_T = a p_T \oplus b$  with $a =2\times 10^{-5}$\,GeV$^{-1}$ and $b = 1\times 10^{-3}$ for both FCC-ee and CEPC~\cite{Dam:2018rfz, CEPCStudyGroup:2018ghi, FCC:2018evy, Bernardi:2022hny}. We ignore the dependence of the momentum resolution on the scattering angle $\theta$.
For convenience, we list $\delta \sqrt{s}/\sqrt{s}$ and $\delta p_T/p_T$ for FCC-ee and CEPC in tables~\ref{table:info_FCC-ee} and~\ref{table:info_CEPC} for the various center-of-mass energies. 

To obtain the background efficiencies $\epsilon_\text{bkg}^{x_c}$ shown in the same tables, we use the muon momentum cut $x_c = 1$ as was done in~\cite{Dam:2018rfz}. This cut gives a signal efficiency of $\epsilon_\text{sig}^{x_c} = 50\%$, while the background is reduced by almost six orders of magnitude $\epsilon_\text{sig}^{x_c} \ll 1$.

In addition, the tables show the expected integrated luminosities $\mathcal L_\text{int}$. Note that CEPC quotes a total integrated luminosity of $100$\,ab$^{-1}$ at the $Z$-pole~\cite{CEPCPhysicsStudyGroup:2022uwl}. We assume that half of that amount will be taken on-peak and the other half slightly off-peak, as is proposed for the FCC-ee~\cite{Bernardi:2022hny}.  

The resulting expected background event numbers $N_\text{bkg}$ are also listed in tables~\ref{table:info_FCC-ee} and~\ref{table:info_CEPC}. We find sizeable numbers of background events on or close to the $Z$-pole, while for the higher energy runs of FCC-ee and CEPC, the searches for $e^+ e^- \to \tau \mu$ have to face very little background. Note that the quoted uncertainties on the background numbers are statistical only. For the $Z$-pole runs, systematic uncertainties might become relevant and could lead to an $\mathcal O(1)$ increase in the background uncertainties. 

The last rows of tables~\ref{table:info_FCC-ee} and~\ref{table:info_CEPC} show our final results for the cross-section sensitivities~$\sigma$. Comparing to the LEP results shown in equations~\eqref{eq:LEP_OPAL1} - \eqref{eq:LEP_OPAL3}, we conclude that FCC-ee and CEPC can improve the sensitivity by more than 3 orders of magnitude.

In the following section, we will translate the cross-section sensitivities into sensitivities to the flavor-changing new physics operators introduced in section~\ref{sec:operators}.

\section{Results and Discussion} \label{sec:results}

Combining the $e^+ e^- \to \tau \mu$ cross-section sensitivities at FCC-ee and CEPC derived in the previous section~\ref{sec:sensitivities} with the expressions for the cross-section derived in section~\ref{sec:dilepton}, we obtain expected constraints on the new physics Wilson coefficients.

In order to statistically combine the expected constraints from $e^+ e^- \to \tau \mu$ at different center-of-mass energies and in order to compare them with low energy constraints, we assume that the existing bounds that we discussed in section~\ref{sec:constraints}, as well as the expected sensitivities that we derived in section~\ref{sec:sensitivities}, correspond to likelihood functions that can be approximated by Gaussians with vanishing central values. 

\begin{figure}[tb]
\centering
\includegraphics[width=0.46\linewidth]{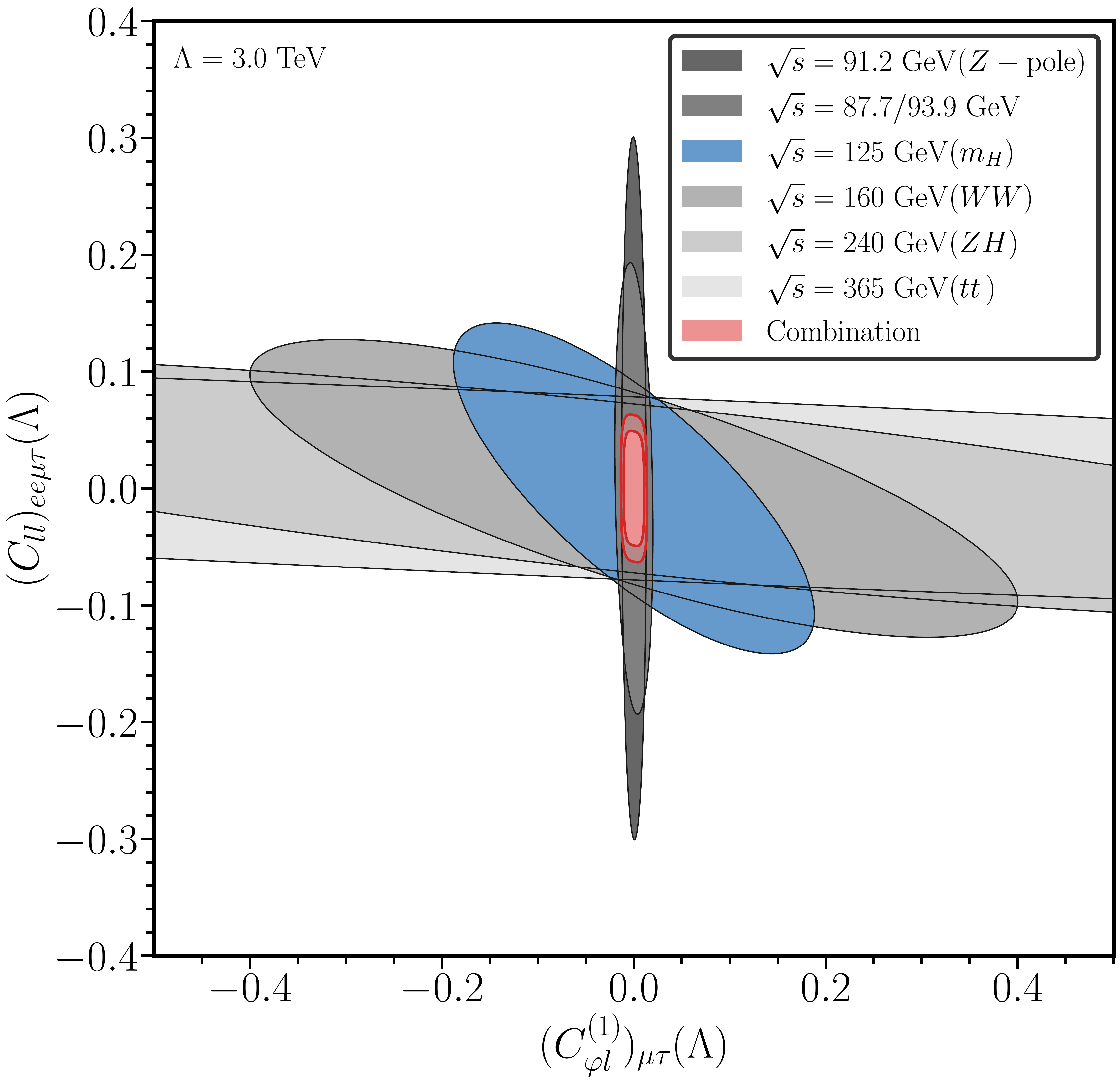} ~~~~
\includegraphics[width=0.46\linewidth]{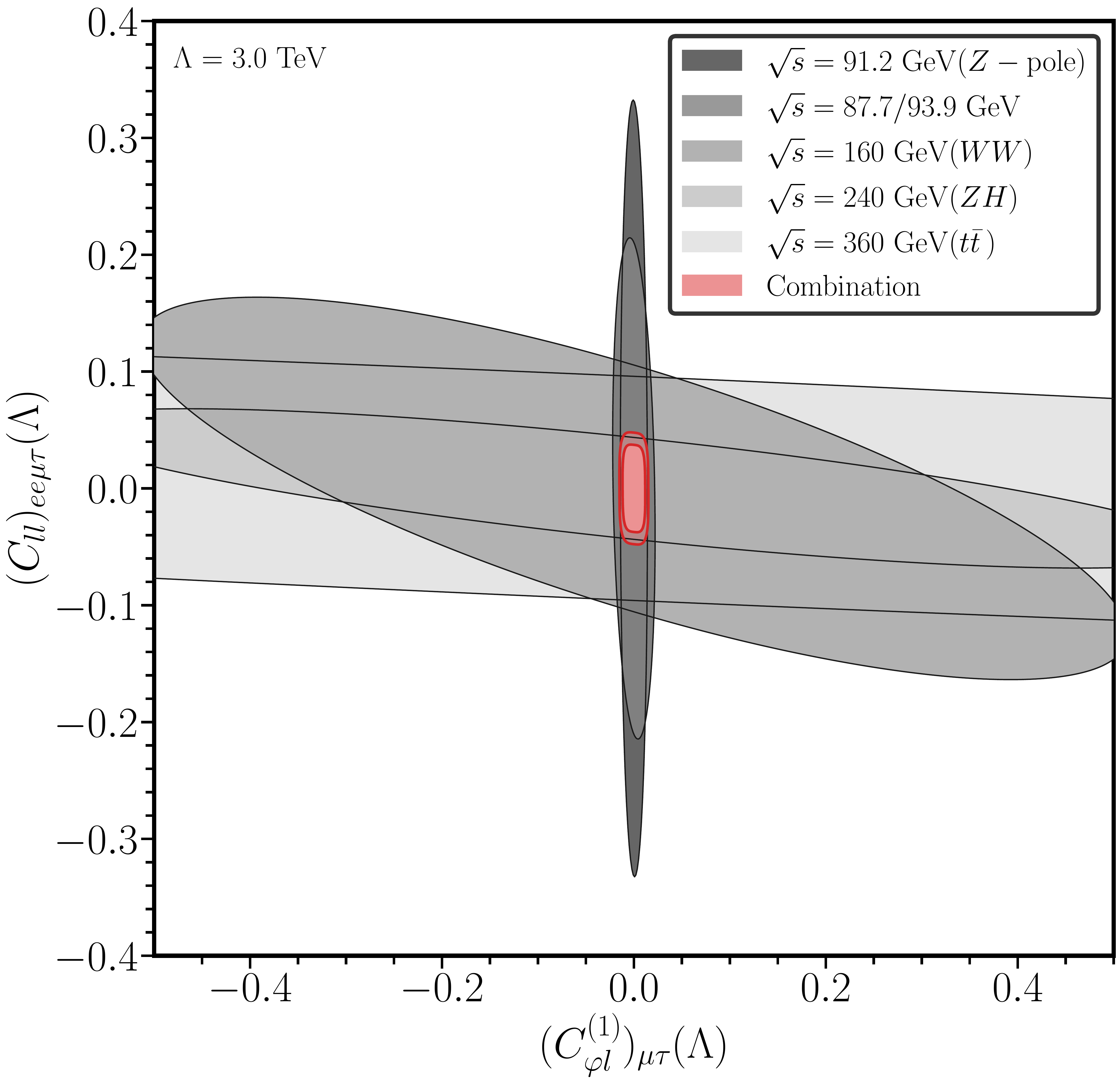}
\caption{Expected constraints in the plane of the SMEFT Wilson coefficients $(C_{ll})_{ee\mu\tau}$ vs. $(C_{\varphi l}^{(1)})_{\mu\tau}$ from $e^+ e^- \to \tau \mu$ searches at FCC-ee (left) and CEPC (right). All other SMEFT coefficients are set to zero, and the new physics scale is set to $\Lambda = 3$~TeV. The various gray regions are the $2\sigma$ constraints from individual runs at the indicated center-of-mass energies. The red regions are the combined $1\sigma$ and $2\sigma$ constraints.} 
\label{fig:Wilson_complementary_1}
\end{figure}
\begin{figure}[tb]
\centering
\includegraphics[width=0.46\linewidth]{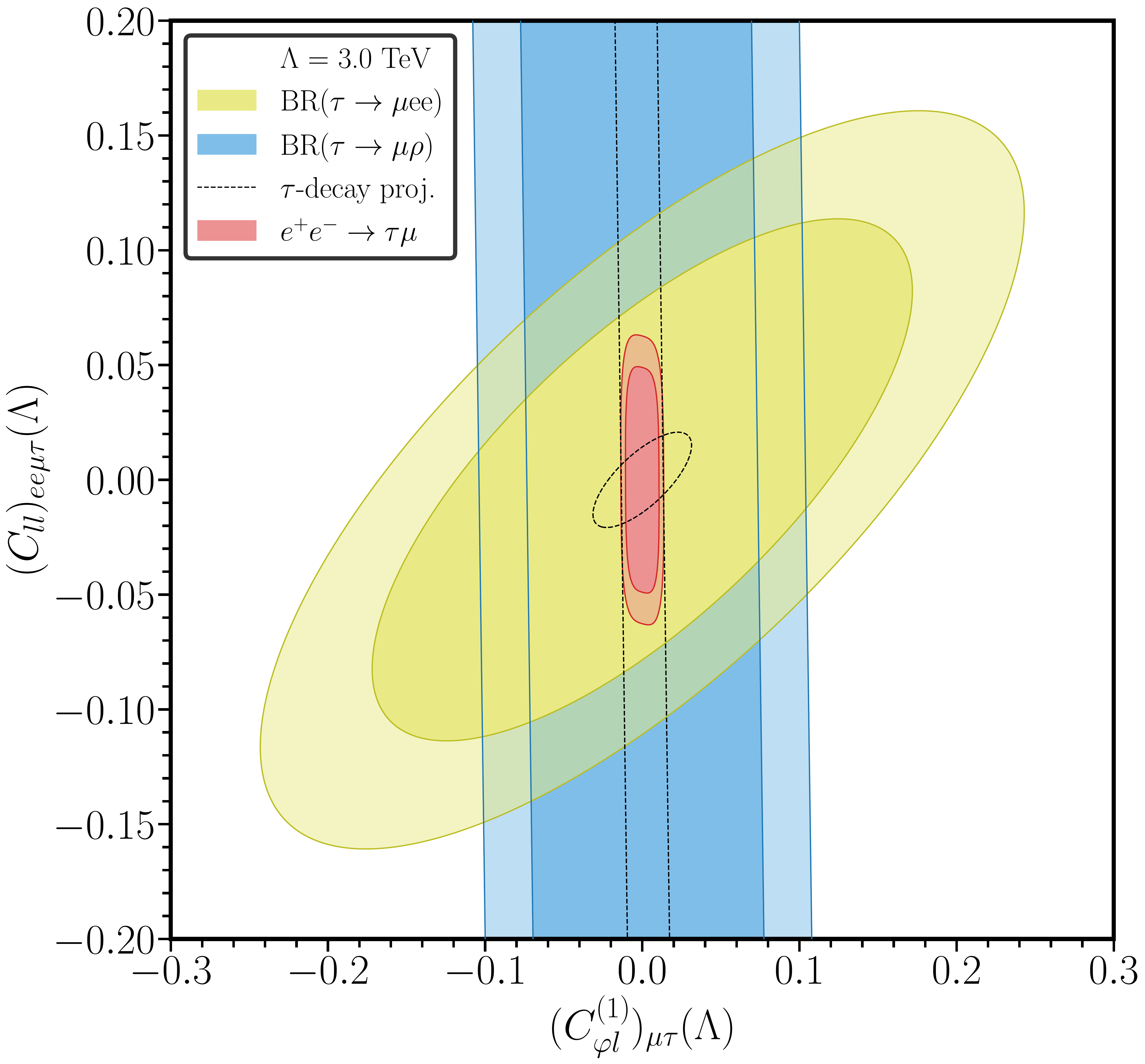} ~~~~
\includegraphics[width=0.46\linewidth]{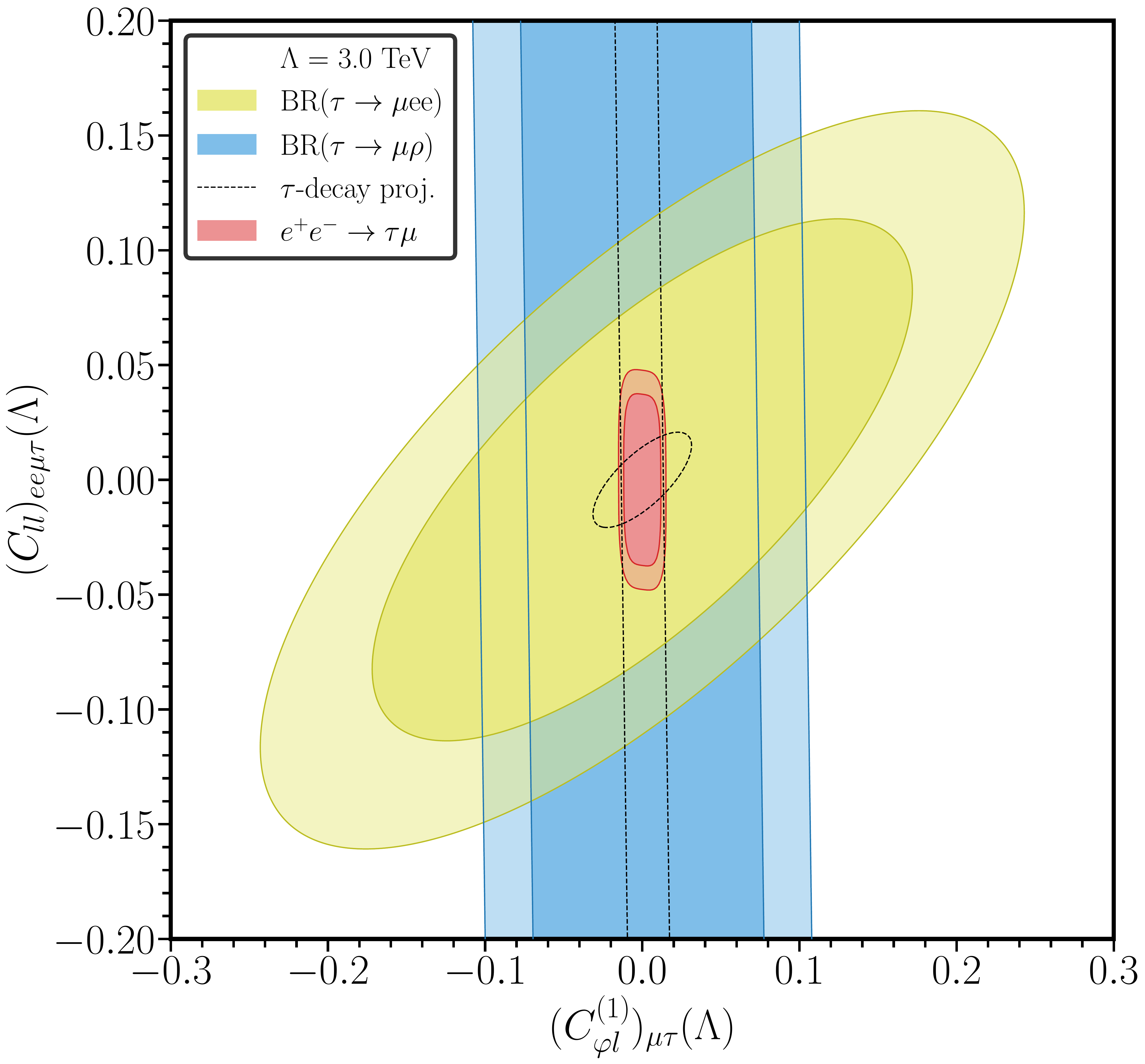}
\caption{Constraints in the plane of the SMEFT Wilson coefficients $(C_{ll})_{ee\mu\tau}$ vs. $(C_{\varphi l}^{(1)})_{\mu\tau}$. All other SMEFT coefficients are set to zero, and the new physics scale is set to $\Lambda = 3$~TeV. The red regions show the expected $1\sigma$ and $2\sigma$ constraints from $e^+ e^- \to \tau \mu$ searches at FCC-ee (left) and CEPC (right) from figure~\ref{fig:Wilson_complementary_1}.
The yellow (blue) region is the current $1\sigma$ and $2\sigma$ constraints from $\tau \to \mu ee$ ($\tau \to \mu \rho$). The black dashed lines show the expected tau decay constraints at $2\sigma$ from Belle II.} 
\label{fig:Wilson_complementary_2}
\end{figure}

In figure~\ref{fig:Wilson_complementary_1}, we show the expected constraints in the plane of the SMEFT Wilson coefficients $(C_{ll})_{ee\mu\tau}$ vs. $(C_{\varphi l}^{(1)})_{\mu\tau}$ from $e^+ e^- \to \tau \mu$ searches at FCC-ee (left) and CEPC (right). All other SMEFT coefficients are set to zero, and the new physics scale is set to $\Lambda = 3$~TeV for definiteness. The gray regions are the $2\sigma$ constraints from individual runs at the indicated center-of-mass energies. The red regions are the combined $1\sigma$ and $2\sigma$ constraints.
Analogous to the results we found from LEP (c.f. figure~\ref{fig:Wilson_current}), the $Z$-pole and near $Z$-pole runs mainly constrain the Wilson coefficient of the Higgs current operator, while the high energy runs mainly constrain the Wilson coefficient of the four-fermion contact operator. Interestingly, a Higgs pole run at FCC-ee (shown in blue) would have comparable sensitivity to both Wilson coefficients. Note the change in scale compared to figure~\ref{fig:Wilson_current}. The expected sensitivities at FCC-ee and CEPC will allow us to probe the Wilson coefficients approximately two orders of magnitude better than at LEP.

Figure~\ref{fig:Wilson_complementary_2} compares the expected sensitivities of FCC-ee and CEPC with the existing and expected sensitivities from rare tau decays. The yellow and blue regions are the current $1\sigma$ and $2\sigma$ constraints from $\tau \to \mu ee$ and $\tau \to \mu \rho$ which give the strongest individual constraints on the considered Wilson coefficients. The black dashed lines show the corresponding expected tau decay constraints at $2\sigma$ from Belle II. We find that the expected sensitivities of FCC-ee and CEPC (shown in red) are much better than the current constraints from the tau decays and comparable to the constraints one can expect from Belle II.

\begin{figure}[tb]
\centering
\includegraphics[width = \linewidth]{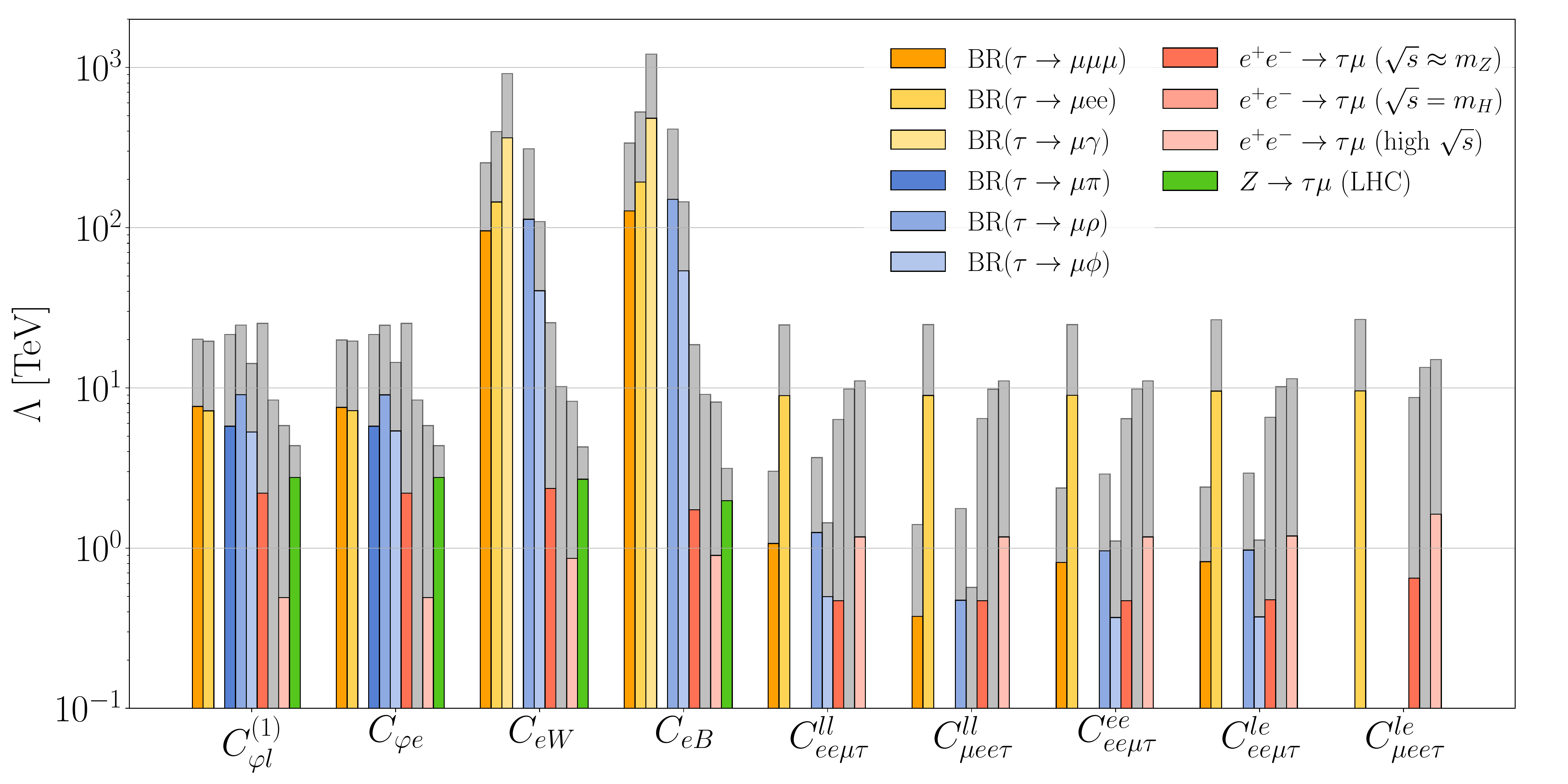}
\caption{Sensitivity to the new physics scale $\Lambda$ from LFV tau decays, $Z$ decays, and the $e^+e^- \to \tau \mu$ cross-section. Each SMEFT Wilson coefficient is turned on individually at the scale $\Lambda$, i.e. $C_i(\Lambda)=1.0$, with all others set to zero. Both current constraints (colored) and future projections (gray) are displayed. The gray bars (FCC-ee sensitivities) on top of the red ones are based on the sensitivities we derived in this work.} 
\label{fig:bar_chart}
\end{figure}

Finally, figure \ref{fig:bar_chart} compares the sensitivity of the various probes to all the lepton flavor violating SMEFT Wilson coefficients we considered. The bar chart shows the new physics scale $\Lambda$ that is excluded by current constraints from BaBar and Belle on leptonic and radiative tau decays (orange) and semi-leptonic tau decays (blue), from the LHC on $Z \to \tau \mu$ decays (green), and from LEP on the $e^+ e^- \to \tau \mu$ cross section (red). Each SMEFT Wilson coefficient is turned on individually at the scale $\Lambda$, i.e. $C_i(\Lambda)=1.0$, with all others set to zero. Currently, the tau decays give the strongest constraint, $\Lambda \gtrsim 10$~TeV for the Higgs current operators and the four-fermion contact operators, and $\Lambda \gtrsim 300$~TeV for the dipole operators.
Expected future sensitivities are shown in gray. 
In the case of the tau decays we show the expectation from Belle II, in the case of the $Z$ decay the expectation from the high-luminosity phase of the LHC. 
In the case of $e^+ e^- \to \tau \mu$ at future machines, we show for each coefficient the expected sensitivity at FCC-ee on the Higgs pole, the best expected sensitivity between the $Z$-pole and near $Z$-pole runs, and the best expected sensitivity between the various high $\sqrt{s}$ runs. The CEPC will have very similar sensitivities.
We observe that searches for $e^+ e^- \to \tau \mu$ at FCC-ee and CEPC will rival the sensitivity of the tau decays.

\section{Conclusions and Outlook} \label{sec:conclusions}

Lepton flavor violation can be searched for in various contexts, e.g. at low energies in tau decays or in the decays of heavy particles like the $Z$ boson or the Higgs. 
In this paper, we studied an alternative probe, the production of $\tau \mu$ at future electron-positron colliders, and determined the expected sensitivity of FCC-ee and CEPC to a $e^+ e^- \to \tau \mu$ signal. 

If new physics is model independently parameterized using SMEFT, searches for $e^+ e^- \to \tau \mu$ give access to three classes of effective operators: flavor violating four-fermion contact interactions, flavor violating dipole interactions, and flavor violating Higgs current interactions. We computed the $e^+ e^- \to \tau \mu$ cross-section in SMEFT and found that the different operators give a signal that shows a very characteristic dependence on the center of mass energy. The contact interactions lead to a $e^+ e^- \to \tau \mu$ cross-section that increases linearly with $s$. The dipole interactions give a cross-section that is constant at large $s$, while the Higgs current interactions lead to a cross-section that falls off like $1/s$ for large $s$. On top of that, the contributions of operators that contain the $Z$ boson are resonantly enhanced on the $Z$ pole. Searches at the different proposed center-of-mass energies at FCC-ee and CEPC thus give complementary information about the operators. In general, we found that FCC-ee and CEPC will be sensitive to new physics scales of $\mathcal O(10~\text{TeV})$, and their sensitivity will rival the sensitivity of searches for lepton flavor violating tau decays at Belle II.

If a $e^+ e^- \to \tau \mu$ signal were to be observed at FCC-ee or CEPC, our study provides a framework to disentangle the contributions from different operators, exploiting the complementarity of searches at the different center-of-mass energies. Additional diagnostics of a signal could be provided by measurements of the forward-backward asymmetry or CP asymmetries. 

The study presented in this paper can be extended in various ways. For example, in addition to $e^+ e^- \to \tau \mu$ one could also consider the processes $e^+e^- \to \tau e$ and $e^+ e^- \to \mu e$. In SMEFT, also t-channel exchange of the photon and the $Z$ boson can contribute to $e^+e^- \to \tau e$, but we expect results that are qualitatively similar to $e^+e^- \to \tau \mu$.
In the case of $e^+e^- \to \mu e$, one expects very strong low energy constraints from lepton flavor violating decays of the muon, like $\mu \to e \gamma$ and $\mu \to 3 e$, as well as $\mu$ to $e$ conversion in nuclei~\cite{Delepine:2001di, Gutsche:2011bi, Calibbi:2021pyh}. 

Another possible extension is to investigate the sensitivity of future linear $e^+ e^-$ machines. As we have seen, the $e^+ e^- \to \tau \mu$ cross-section that is induced by four-fermion contact interactions grows with the center-of-mass energy. The large $\sqrt{s}$ that could be achieved at ILC~\cite{ILC:2013jhg, ILCInternationalDevelopmentTeam:2022izu}, C$^3$~\cite{Bai:2021rdg}, or CLIC~\cite{CLICdp:2018cto, Brunner:2022usy} could thus give unique sensitivity to such operators, see e.g.~\cite{Murakami:2014tna, Cho:2018mro, Etesami:2021hex}. Moreover, the beam polarization at linear colliders could be exploited to  
gain information about the chirality structure of the lepton flavor violating operators.

\section*{Acknowledgements} 

We thank Aneesh Manohar for valuable clarifications about the SMEFT renormalization group evolution and Andreas Crivellin for discussions on lepton flavor violating $Z$ decays. The Feynman diagrams shown in this paper were created using \texttt{TikZ-Feynman}~\cite{Ellis:2016jkw}. The research of W.A. and P.M. is supported by the U.S. Department of Energy grant number DE-SC0010107.

\begin{appendix}
\section{Lepton Flavor Violating Tau Decays} \label{app:tau_decays}

In this appendix, we detail our treatment of the lepton flavor violating tau decays that we use to constrain the new physics Wilson coefficients. 
The appropriate framework to describe the tau decays is the so-called Low-Energy Effective Field Theory (LEFT), where the heavy SM states ($W^\pm$ bosons, $Z$ boson, top quark, and Higgs boson) have been integrated out~\cite{Jenkins:2017jig, Jenkins:2017dyc}.
The list of relevant low-energy operators consists of lepton flavor violating photon dipoles and four fermion contact interactions
\begin{align} \label{eq:tau_op_start}
    (L_\gamma^{LR})_{\mu\tau} & \frac{1}{\sqrt{2}} \frac{v}{\Lambda^2} (\bar \mu \sigma^{\alpha\beta} P_R\tau) F_{\alpha\beta} ~, \quad & 
    (L_\gamma^{RL})_{ee\mu\tau} & \frac{1}{\sqrt{2}} \frac{v}{\Lambda^2} (\bar \mu \sigma^{\alpha\beta} P_L \tau) F_{\alpha\beta} ~, \\[16pt]
    (L_{V,e}^{LL})_{\mu\tau} & \frac{1}{\Lambda^2} (\bar e \gamma_\alpha P_L e)(\bar \mu \gamma^{\alpha} P_L \tau) ~, \quad & (L_{V,e}^{RR})_{\mu\tau} & \frac{1}{\Lambda^2} (\bar e \gamma_\alpha P_R e)(\bar \mu \gamma^{\alpha} P_R \tau) ~, \\
    (L_{V,e}^{LR})_{\mu\tau} & \frac{1}{\Lambda^2} (\bar e \gamma_\alpha P_L e)(\bar \mu \gamma^{\alpha} P_R \tau) ~, \quad & (L_{V,e}^{RL})_{\mu\tau} & \frac{1}{\Lambda^2} (\bar e \gamma_\alpha P_R e)(\bar \mu \gamma^{\alpha} P_L \tau) ~, \\
    (L_{S,e}^{LR})_{\mu\tau} & \frac{1}{\Lambda^2} (\bar e P_L e)(\bar \mu P_R \tau) ~, \quad & (L_{S,e}^{RL})_{\mu\tau} & \frac{1}{\Lambda^2} (\bar e P_R e)(\bar \mu P_L \tau) ~, \\[16pt]
    (L_{V,\mu}^{LL})_{\mu\tau} & \frac{1}{\Lambda^2} (\bar \mu \gamma_\alpha P_L \mu)(\bar \mu \gamma^{\alpha} P_L \tau) ~, \quad & (L_{V,\mu}^{RR})_{\mu\tau} & \frac{1}{\Lambda^2} (\bar \mu \gamma_\alpha P_R \mu)(\bar \mu \gamma^{\alpha} P_R \tau) ~, \\
    (L_{V,\mu}^{LR})_{\mu\tau} & \frac{1}{\Lambda^2} (\bar \mu \gamma_\alpha P_L \mu)(\bar \mu \gamma^{\alpha} P_R \tau) ~, \quad & (L_{V,\mu}^{RL})_{\mu\tau} & \frac{1}{\Lambda^2} (\bar \mu \gamma_\alpha P_R \mu)(\bar \mu \gamma^{\alpha} P_L \tau) ~, \\[16pt]
    (L_{V,q}^{LL})_{\mu\tau} & \frac{1}{\Lambda^2} (\bar q \gamma_\alpha P_L q)(\bar \mu \gamma^{\alpha} P_L \tau) ~, \quad & (L_{V,q}^{RR})_{\mu\tau} & \frac{1}{\Lambda^2} (\bar q \gamma_\alpha P_R q)(\bar \mu \gamma^{\alpha} P_R \tau) ~, \\
    (L_{V,q}^{LR})_{\mu\tau} & \frac{1}{\Lambda^2} (\bar q \gamma_\alpha P_L q)(\bar \mu \gamma^{\alpha} P_R \tau) ~, \quad & (L_{V,q}^{RL})_{\mu\tau} & \frac{1}{\Lambda^2} (\bar q \gamma_\alpha P_R q)(\bar \mu \gamma^{\alpha} P_L \tau) ~, \label{eq:tau_op_end}
\end{align}
where $q$ denotes the light quarks $q = u, d, s$. 

The tree-level matching conditions between the SMEFT Wilson coefficients that we use to describe $e^+ e^- \to \tau \mu$ and the LEFT coefficients can be determined from the results in~\cite{Jenkins:2017jig} and are given by
\begin{align}
(L_\gamma^{LR})_{\mu\tau} & = (C_\gamma^{LR})_{\mu\tau} ~,  & (L_\gamma^{RL})_{\mu\tau} & = (C_\gamma^{RL})_{\mu\tau} ~, \\[16pt]
(L_{V,e}^{LL})_{\mu\tau} & = (C_V^{LL})_{\mu \tau} - (1-2s_W^2) (C_Z^{LL})_{\mu \tau} ~,  & (L_{V,e}^{RR})_{\mu\tau} & = (C_V^{RR})_{\mu \tau} + 2s_W^2 (C_Z^{RR})_{\mu \tau} ~, \\
(L_{V,e}^{LR})_{\mu\tau} & = (C_V^{LR})_{\mu \tau} - (1-2s_W^2) (C_Z^{RR})_{\mu \tau} ~,  & (L_{V,e}^{RL})_{\mu\tau} & = (C_V^{RL})_{\mu \tau} + 2s_W^2 (C_Z^{LL})_{\mu \tau} ~, \\
(L_{S,e}^{LR})_{\mu\tau} & = (C_S^{LR})_{\mu\tau} ~,  & (L_{S,e}^{RL})_{\mu\tau} & = (C_S^{RL})_{\mu\tau} ~, \\[16pt]
(L_{V,\mu}^{LL})_{\mu\tau} & = - (1-2s_W^2) (C_Z^{LL})_{\mu \tau} ~,  & (L_{V,\mu}^{RR})_{\mu\tau} & = 2s_W^2 (C_Z^{RR})_{\mu \tau} ~, \\
(L_{V,\mu}^{LR})_{\mu\tau} & = - (1-2s_W^2) (C_Z^{RR})_{\mu \tau} ~,  & (L_{V,\mu}^{RL})_{\mu\tau} & = 2s_W^2 (C_Z^{LL})_{\mu \tau} ~, \\[16pt]
(L_{V,u}^{LL})_{\mu\tau} & = (1-\frac{4}{3} s_W^2) (C_Z^{LL})_{\mu \tau} ~,  & (L_{V,u}^{RR})_{\mu\tau} & = -\frac{4}{3} s_W^2 (C_Z^{RR})_{\mu \tau} ~, \\
(L_{V,u}^{LR})_{\mu\tau} & = (1-\frac{4}{3} s_W^2) (C_Z^{RR})_{\mu \tau} ~,  & (L_{V,u}^{RL})_{\mu\tau} & = -\frac{4}{3} s_W^2 (C_Z^{LL})_{\mu \tau} ~, \\[16pt]
(L_{V,d}^{LL})_{\mu\tau} & = (L_{V,s}^{LL})_{\mu\tau} = -(1-\frac{2}{3} s_W^2) (C_Z^{LL})_{\mu \tau} ~,  & (L_{V,d}^{RR})_{\mu\tau} & = (L_{V,s}^{RR})_{\mu\tau} = \frac{2}{3} s_W^2 (C_Z^{RR})_{\mu \tau} ~, \\
(L_{V,d}^{LR})_{\mu\tau} & = (L_{V,s}^{LR})_{\mu\tau} = -(1-\frac{2}{3} s_W^2) (C_Z^{RR})_{\mu \tau} ~,  & (L_{V,d}^{RL})_{\mu\tau} & = (L_{V,s}^{RL})_{\mu\tau} = \frac{2}{3} s_W^2 (C_Z^{LL})_{\mu \tau} ~.
\end{align}
Going beyond leading order, one needs to (i) run the SMEFT coefficients from the new physics scale $\Lambda$ to the electroweak scale $m_Z$, (ii) match the SMEFT coefficients onto the relevant LEFT operator coefficients at the scale $m_Z$, and (iii) run the LEFT coefficients from the electroweak scale to the low-energy scale $m_\tau$.

In our numerical analysis, we incorporate the 1-loop running in SMEFT due to electroweak gauge couplings and the top Yukawa coupling, implementing the results from~\cite{Jenkins:2013wua, Alonso:2013hga}, and 1-loop running in LEFT due to QED, implementing the results from~\cite{Jenkins:2017dyc}. Throughout, we work in the leading logarithmic approximation. Note that beyond leading order, one needs to extend the set of operators given in eqs.~\eqref{eq:tau_op_start}-\eqref{eq:tau_op_end} by four fermion operators that contain heavy quark or tau pairs, $b \bar b$, $c \bar c$, or $\tau^+ \tau^-$. Such operators are generated in the matching~\cite{Jenkins:2017jig} and mix at 1-loop into the operators relevant for the tau decays~\cite{Jenkins:2017dyc}. Numerically, we find that going beyond leading order gives only percent-level corrections.

\begin{figure}[tb]
\centering
\includegraphics[width=0.28\linewidth]{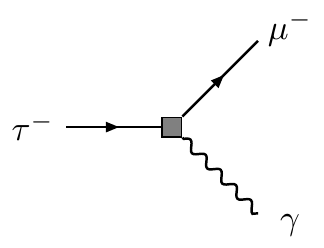} ~~
\includegraphics[width=0.35\linewidth]{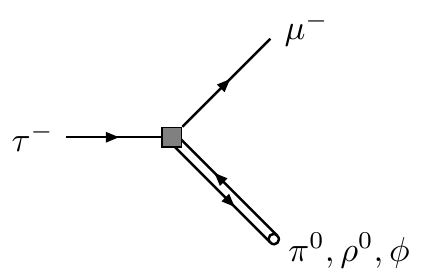} ~~ 
\includegraphics[width=0.31\linewidth]{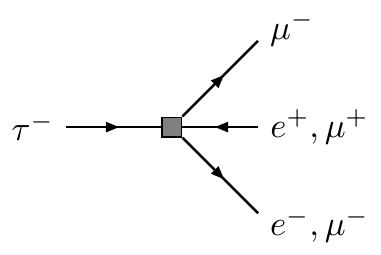} \\[16pt]
\includegraphics[width=0.42\linewidth]{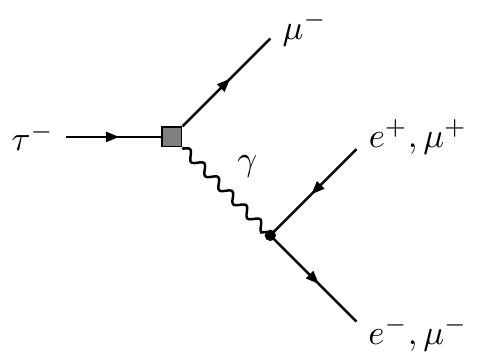} \qquad 
\includegraphics[width=0.42\linewidth]{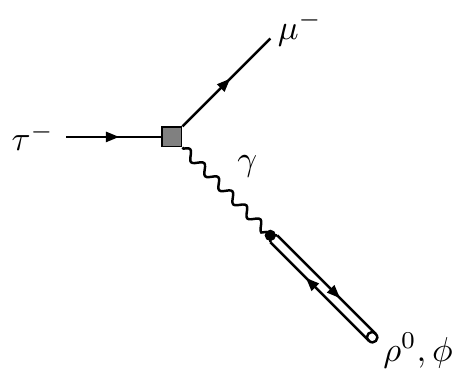} 
\caption{Feynman diagrams for the lepton flavor violating tau decays $\tau^- \to \mu^- \gamma$, $\tau^- \to \mu^- e^+e^-$, $\tau^- \to \mu^- \mu^+ \mu^-$, $\tau^- \to \mu^- \pi^0$, $\tau^- \to \mu^- \rho^0$, and $\tau^- \to \mu^- \phi$.}
\label{fig:diagrams_tau_decays}
\end{figure}

The branching ratios of flavor-changing tau decays can be expressed in terms of the LEFT Wilson coefficients at the low scale $m_\tau$. We will focus on the tau decays that give the strongest constraints, namely the radiative decay $\tau\to \mu \gamma$, the leptonic decays $\tau \to 3 \mu$, $\tau \to \mu e^+ e^-$, and the semi-leptonic decays $\tau \to \mu \pi$, $\tau \to \mu \rho$, $\tau \to \mu \phi$. The relevant Feynman diagrams are shown in figure~\ref{fig:diagrams_tau_decays}.

To simplify the expressions, we find it convenient to normalize the lepton flavor violating branching ratios to appropriate branching ratios of well-known SM tau decays like $\tau \to \mu \nu \nu$, $\tau \to \nu \pi$, and $\tau \to \nu \rho$. We find
\begin{multline}
\frac{\text{BR}(\tau^- \to \mu^- e^+ e^-)}{\text{BR}(\tau^- \to \mu^- \nu_\tau \bar\nu_\mu)} \simeq \frac{v^4}{4 \Lambda^4} \Bigg[ |L_{V,e}^{LL}|^2 + |L_{V,e}^{RR}|^2 + |L_{V,e}^{LR}|^2 + |L_{V,e}^{RL}|^2 + \frac{1}{4} |L_{S,e}^{LR}|^2 + \frac{1}{4} |L_{S,e}^{LR}|^2 \\
+ 16 e^2 \frac{v^2}{m_\tau^2} \Big( |L_\gamma^{LR}|^2 + |L_\gamma^{RL}|^2 \Big) \left( \log\left(\frac{m_\tau^2}{m_e^2}\right) - 3 \right) \\
- 4 \sqrt{2} e \frac{v}{m_\tau} \text{Re} \Big( L_\gamma^{LR} (L_{V,e}^{LL} + L_{V,e}^{RL})^* + L_\gamma^{RL} (L_{V,e}^{LR} + L_{V,e}^{RR})^* \Big) \Bigg] ~,
\end{multline}
\begin{multline}
\frac{\text{BR}(\tau^- \to \mu^- \mu^+ \mu^-)}{\text{BR}(\tau^- \to \mu^- \nu_\tau \bar\nu_\mu)} \simeq \frac{v^4}{4 \Lambda^4} \Bigg[ 2|L_{V,\mu}^{LL}|^2 + 2|L_{V,\mu}^{RR}|^2 + |L_{V,\mu}^{LR}|^2 + |L_{V,\mu}^{RL}|^2  \\
+ 16 e^2 \frac{v^2}{m_\tau^2} \Big( |L_\gamma^{LR}|^2 + |L_\gamma^{RL}|^2 \Big) \left( \log\left(\frac{m_\tau^2}{m_e^2}\right) - \frac{11}{4} \right) \\
- 4 \sqrt{2} e \frac{v}{m_\tau} \text{Re} \Big( L_\gamma^{LR} (2L_{V,\mu}^{LL} + L_{V,\mu}^{RL})^* + L_\gamma^{RL} (2L_{V,\mu}^{LR} + L_{V,\mu}^{RR})^* \Big) \Bigg] ~,
\end{multline}
\begin{equation}
\frac{\text{BR}(\tau^- \to \mu^- \gamma)}{\text{BR}(\tau^- \to \mu^- \nu_\tau \bar\nu_\mu)} \simeq 48 \pi^2 \frac{v^6}{m_\tau^2 \Lambda^4} \Big( |L_\gamma^{LR}|^2 + |L_\gamma^{RL}|^2  \Big) ~,
\end{equation}
\begin{equation}
\frac{\text{BR}(\tau^- \to \mu^- \pi^0)}{\text{BR}(\tau^- \to \nu \pi^-)} \simeq \frac{v^4}{8 \Lambda^4} \Big( |L_{V,u}^{LL} - L_{V,d}^{LL} - L_{V,u}^{RL} + L_{V,d}^{RL}|^2 + |L_{V,u}^{RR} - L_{V,d}^{RR} - L_{V,u}^{LR} + L_{V,d}^{LR} |^2  \Big) ~,
\end{equation}
\begin{multline}
\frac{\text{BR}(\tau^- \to \mu^- \rho^0)}{\text{BR}(\tau^- \to \nu \rho^-)} \simeq \frac{v^4}{8 \Lambda^4} \left(1 + \frac{2m_\rho^2}{m_\tau^2}\right)^{-1} \Bigg[ 8 e^2 \frac{v^2}{m_\rho^2} \Big( |L_\gamma^{LR}|^2 + |L_\gamma^{RL}|^2 \Big) \left(2+ \frac{m_\rho^2}{m_\tau^2} \right) \\
+ 12 \sqrt{2} e \frac{v}{m_\tau} \text{Re}\Big( L_\gamma^{LR}(L_{V,u}^{LL} - L_{V,d}^{LL} + L_{V,u}^{RL} - L_{V,d}^{RL})^* + L_\gamma^{RL} (L_{V,u}^{RR} - L_{V,d}^{RR} + L_{V,u}^{LR} - L_{V,d}^{LR})^* \Big) \\ 
+ \Big( |L_{V,u}^{LL} - L_{V,d}^{LL} + L_{V,u}^{RL} - L_{V,d}^{RL}|^2 + |L_{V,u}^{RR} - L_{V,d}^{RR} + L_{V,u}^{LR} - L_{V,d}^{LR} |^2  \Big) \left(1 + \frac{2m_\rho^2}{m_\tau^2}\right) \Bigg] ~,
\end{multline}
\begin{multline}
\frac{\text{BR}(\tau^- \to \mu^- \phi)}{\text{BR}(\tau^- \to \nu \rho^-)} \simeq \frac{v^4}{4 \Lambda^4} \frac{f_\phi^2}{f_\rho^2} \left(1 + \frac{2m_\rho^2}{m_\tau^2}\right)^{-1} \left(1 - \frac{m_\rho^2}{m_\tau^2}\right)^{-2}\left(1 - \frac{m_\phi^2}{m_\tau^2}\right)^{2} \\
\times \Bigg[ \frac{8}{9} e^2 \frac{v^2}{m_\phi^2} \Big( |L_\gamma^{LR}|^2 + |L_\gamma^{RL}|^2 \Big) \left(2+ \frac{m_\phi^2}{m_\tau^2} \right) + \Big( |L_{V,s}^{LL} + L_{V,s}^{RL}|^2 + |L_{V,s}^{RR} + L_{V,s}^{LR}|^2  \Big) \left(1 + \frac{2m_\phi^2}{m_\tau^2}\right) \\
- 4 \sqrt{2} e \frac{v}{m_\tau} \text{Re}\Big( L_\gamma^{LR}(L_{V,s}^{LL} + L_{V,s}^{RL})^* + L_\gamma^{RL}(L_{V,s}^{RR} + L_{V,s}^{LR})^* \Big) \Bigg] ~,
\end{multline}
where for better readability we omitted the flavor subscript ``$\mu\tau$'' on the Wilson coefficients. To obtain numerical predictions for the lepton flavor violating tau decays, one can use the PDG values~\cite{ParticleDataGroup:2022pth}: $\text{BR}(\tau^- \to \mu^- \nu_\tau \bar\nu_\mu) \simeq 17.39\,\%$, $\text{BR}(\tau^- \to \nu \pi^-) \simeq 10.82\,\%$, and $\text{BR}(\tau^- \to \nu \rho^-) \simeq 25.49\,\%$. The ratio of $\phi$ and $\rho$ decay constants can be extracted from~\cite{Bharucha:2015bzk}, $f_\phi/f_\rho \simeq 1.09$.

\end{appendix}


\bibliography{bibliography}

\end{document}